\documentclass[12pt,preprint]{aastex}

\shorttitle{The early X-ray emission from GRBs}
\shortauthors{Paul T. O'Brien et al.}

\begin{document}
\def\grb{{GRB050502b}}
\def\swift{{\it Swift}}
\def\lsun{{\rm L_{\odot}}}
\def\msun{{\rm M_{\odot}}}
\def\rsun{{\rm R_{\odot}}}
\def\go{
\mathrel{\raise.3ex\hbox{$>$}\mkern-14mu\lower0.6ex\hbox{$\sim$}}
}
\def\lo{
\mathrel{\raise.3ex\hbox{$<$}\mkern-14mu\lower0.6ex\hbox{$\sim$}}
}
\def\simeq{
\mathrel{\raise.3ex\hbox{$\sim$}\mkern-14mu\lower0.4ex\hbox{$-$}}
}

\title{The early X-ray emission from GRBs}
\author{
P.T. O'Brien\altaffilmark{1},
R. Willingale\altaffilmark{1},
J. Osborne\altaffilmark{1},
M.R. Goad\altaffilmark{1},
K.L. Page\altaffilmark{1},
S. Vaughan\altaffilmark{1},
E. Rol\altaffilmark{1}, 
A. Beardmore\altaffilmark{1},
O. Godet\altaffilmark{1}, 
C.P. Hurkett\altaffilmark{1},
A. Wells\altaffilmark{1}, 
B. Zhang\altaffilmark{2}, 
S. Kobayashi\altaffilmark{3}, 
D.N. Burrows\altaffilmark{4}, 
J.A. Nousek\altaffilmark{4},
J.A. Kennea\altaffilmark{4},
A. Falcone\altaffilmark{4},
D. Grupe\altaffilmark{4}, 
N. Gehrels\altaffilmark{5}, 
S. Barthelmy\altaffilmark{5},
J. Cannizzo\altaffilmark{5,6},
J. Cummings\altaffilmark{5,7}, 
J.E. Hill\altaffilmark{5,8},
H. Krimm\altaffilmark{5,8},
G. Chincarini\altaffilmark{9,10}, 
G. Tagliaferri\altaffilmark{9},
S. Campana\altaffilmark{9}, 
A. Moretti\altaffilmark{9}, 
P. Giommi\altaffilmark{11}, 
M. Perri\altaffilmark{11}, 
V. Mangano\altaffilmark{12},
V. LaParola\altaffilmark{12}
}
\altaffiltext{1}{Department of Physics \& Astronomy,
University of Leicester, Leicester LE1 7RH, UK}
\altaffiltext{2}{Department of Physics, University of Nevada, Las Vegas, 
NV 89154}
\altaffiltext{3}{Astrophysics Research Institute, Liverpool John Moores 
University, Twelve Quays House, Birkenhead CH41 1LD, UK}
\altaffiltext{4}{Department of Astronomy and Astrophysics, Pennsylvania
State University, 525 Davey Laboratory, University Park, PA 16802}
\altaffiltext{5}{NASA/Goddard Space Flight Center, Greenbelt, MD 20771}
\altaffiltext{6}{Joint Center for Astrophysics, University of Maryland,
Baltimore County, Baltimore, MD 21250}
\altaffiltext{7}{National Research Council, 2101 Constitution Avenue,
Washington, DC 20418}
\altaffiltext{8}{Universities Space Research Association, 10211 
Wincopin Circle,
Suite 500, Columbia, MD 21044-3432}
\altaffiltext{9}{INAF, Osservatorio Astronomico di Brera, Via E. Bianchi 46,
I-23807 Merate (LC), Italy}
\altaffiltext{10}{Universit\`a degli Studi di Milano-Bicocca, Dipartimento 
di Fisica, Piazza delle Scienze 3, I-20126 Milano, Italy}
\altaffiltext{11}{ASI Science Data Center, via G. Galilei, I-00044
Frascati, Italy}
\altaffiltext{12}{INAF, istituto di Astrofisica Spaziale e Fisica Cosmica
Sezione di Palermo, Via U. La Malfa 153, I-90146 Palermo, Italy}

\begin{abstract} 
We present observations of the early X-ray emission for a sample of 40
gamma--ray bursts (GRBs) obtained using the \swift\ satellite for
which the narrow-field instruments were pointed at the burst within 10
minutes of the trigger. Using data from the Burst Alert and X-Ray
Telescopes, we show that the X-ray light curve can be well described
by an exponential that relaxes into a power law, often with flares
superimposed. The transition time between the exponential and the power
law provides a physically defined timescale for the burst duration.
In most bursts the power law breaks to a shallower decay within the
first hour, and a late emission ``hump'' is observed which can last
for many hours. In other GRBs the hump is weak or absent.  The
observed variety in the shape of the early X-ray light curve can be
explained as a combination of three components: prompt emission from
the central engine; afterglow; and the late hump. In this scenario,
afterglow emission begins during or soon after the burst and the
observed shape of the X-ray light curve depends on the relative
strengths of the emission due to the central engine and that of the
afterglow.  There is a strong correlation such that those GRBs with
stronger afterglow components have brighter early optical emission.
The late emission hump can have a total fluence equivalent to that of
the prompt phase. GRBs with the strongest late humps have weak or no
X-ray flares.

\end{abstract}

\section{Introduction}

Gamma-ray bursts (GRBs) are identified as a brief flash of gamma-rays
seen at a random location on the sky. For that instant the GRB becomes
the intrinsically brightest single object in the Universe.  The
duration of a GRB in terms of its prompt emission (i.e., the
``burst'') is usually defined in terms of the timescale over which
90\% of the gamma-rays were detected -- the T$_{90}$ parameter.  It is
conventional to describe those GRBs for which T$_{90}$ $> 2$s as
``long/soft bursts'' and those with shorter duration as ``short/hard
bursts'' (e.g. Kouveliotou et al. 1993).

It is now generally accepted that long-duration GRBs result from the
death of a rapidly-rotating massive star (Paczy\'nski, 1998;
MacFadyen \& Woosley, 1999).  The stellar core collapses inwards to
form a black hole surrounded by an accreting disk or torus.  The
accreting material liberates gravitational potential energy either in
the form of neutrinos or via magneto-hydrodynamic processes. These
generate a relativistic jet, oriented along the rotation axis of the
stellar core, which eventually escapes the star. The jet contains a
relatively modest amount of baryonic material moving at high Lorentz
factor.
 
Prior to the \swift\ era, almost all of our knowledge about the
emission from GRBs beyond a few seconds came from the study of long
bursts. Short bursts were, quite literally, too short to be localized
by observatories prior to \swift . The \swift\ satellite (Gehrels et
al. 2004) has changed that situation. The data for the first short
bursts detected by \swift\ (Gehrels et al. 2005; Barthelmy et
al. 2005b; Hjorth et al. 2005; Bloom et al. 2006) strongly support the
idea that the gamma-ray emission seen from short GRBs arises from a
jet powered by a merger of two compact objects, most likely two
neutron stars or a neutron star and a black hole. The
\swift\ data also show that the X-ray emission from some short bursts
can be detected long after T$_{90}$, allowing for a direct comparison
of the early X-ray emission between short and long GRBs. Some short
bursts may have collimated flow (e.g. GRB051221A, Burrows et al. 2006)
although this is less clear in others (e.g. GRB050724, Barthelmy et
al. 2005b).

For both types of GRB, it is thought that the jet flow is
inhomogeneous, leading to internal shocks caused by a variable Lorentz
factor (Rees \& M\'esz\'aros 1994; Sari \& Piran 1997). These produce
the variable gamma-ray emission seen, when viewing within the jet
beam, as a GRB.  The gamma-ray emission typically lasts a few tens of
seconds before fading below detectability with the current generation
of gamma-ray instruments, including the \swift\ Burst Alert Telescope
(BAT; Barthelmy et al. 2005a).  The sensitivity of BAT is a
complicated function of burst duration and spectral shape (see Band et
al. 2006). In practice the BAT detects bursts with 15--150 keV
fluences as low as $\sim 10^{-8}$ erg cm$^{-2}$. The \swift\
satellite, however, has the capability to rapidly slew and point its
X-ray Telescope (XRT; Burrows et al. 2005b) and Ultraviolet Optical
Telescope (UVOT; Roming et al. 2005) at the burst location. The XRT
can detect a source at fainter flux levels in the (observed) 0.3--10
keV band than are possible using extrapolated BAT data. For a
Crab-like spectrum and a Galactic column of $1\times 10^{20}$
cm$^{-2}$ the XRT detects $\sim 1$ count per 100s for a source with a
0.3--10 keV flux of $10^{-12}$ erg cm$^{-2}$ s$^{-1}$. Thus, \swift\
can routinely follow the evolution of the earliest X-ray emission from
GRBs with only modest gaps in the observed light curve.

The GRB flux can be represented as a function of time and frequency
using a function $f_\nu\propto \nu^{-\beta} t^{-\alpha}$, where
$\beta$ is the spectral index and $\alpha$ is the temporal
index\footnote{The photon index $\Gamma$ is related to $\beta$ by
$\Gamma = \beta + 1$.}.  Analysis of some of the first bursts observed
with \swift\ shows that the early X-ray light curves are complex. In
some cases (e.g. Tagliaferri et al. 2005; Hill et al. 2006; Cusumano
et al. 2006; Vaughan et al. 2006a) the early X-ray emission (observed
with the XRT within a few hundred seconds of the burst) can decline
rapidly in the first few minutes, with $\alpha \sim 3$ or greater. The
light curve can then show a break to a shallower decay component,
which we will henceforth refer to as the late emission
``hump''. Analysis of GRB samples from the early phase of the \swift\
mission (Nousek et al. 2006) confirm that this pattern of a rapidly
decaying light curve followed by a late emission hump is common.
However, in some bursts the earliest observed X-ray flux appears to
decline relatively slowly ($\alpha \sim 1$) (e.g. Campana et al. 2005;
Blustin et al. 2006).

The fading X-ray emission could be due to a number of components,
including high-latitude emission from the fading burst (Kumar \&
Panaitescu 2000), the interaction of the jet with the surroundings ---
the afterglow emission produced by an external shock
(e.g. M\'esz\'aros \& Rees 1997), and thermal emission from a
photosphere around the outflow (e.g. M\'esz\'aros \& Rees 2000) or
from a hot cocoon associated with the jet (e.g. M\'esz\'aros \& Rees
2001). A significant fraction of GRBs also show X-ray flares
(e.g. Burrows et al. 2005a) superimposed on the declining light
curves.

If the BAT and the XRT are initially detecting the prompt emission
from the jet, the most rapidly-decaying X-ray light curves could be
due to viewing photons at high-latitudes (i.e. large angles to the
line-of-sight) as the prompt emission fades (Kumar \& Panaitescu 2000;
Tagliaferri et al. 2005; Nousek et al. 2006; Zhang et al., 2006;
Panaitescu et al. 2006). In at least one case (GRB050219a, Tagliaferri
et al. 2005), the BAT and XRT light curves do not appear to join and in other
cases the observed X-ray rapid decay rate is higher than expected from
the high-latitude model (e.g. Vaughan et al. 2006a). Alternatives such
as structured jets (e.g.  Zhang \& M\'esz\'aros 2002), multiple jets
or patchy jets (e.g. Burrows et al. 2005c) in which we see varying
emission as our light cone expands are also possibilities.  These
models have difficulties, however, explaining those bursts that
decline relatively slowly for which other emission components, such as
the afterglow, may contribute at the earliest times.  Both the
rapid-decay and afterglow models have difficulties explaining the late
emission hump.  This component may be due to forward shock emission,
which is refreshed with energy either due to continued emission from
the central engine or because the ejecta have a range in initial
Lorentz factor (Rees \& M\'esz\'aros 1998; Sari \& M\'esz\'aros 2000;
Zhang \& M\'esz\'aros 2001; Nousek et al. 2006; Zhang et al. 2006;
Granot \& Kumar 2006).

To disentangle the relative contribution of emission from the central
engine and that due to afterglow, and hence understand the origin of
the early gamma-ray and X-ray emission, requires a systematic analysis
of the temporal and spectral properties of a large GRB sample
combining data from the BAT and XRT.  The previous studies of GRB
samples from \swift\ (e.g. Nousek et al. 2006) include a relatively
small number of GRBs with early (few minutes from trigger) XRT
observations and do not include a combined temporal and spectral
analysis of data from the BAT and XRT.  The aim of this paper is to
determine the shape of the early X-ray light curve and spectrum for a
large sample of GRBs for which observations were obtained early with
the \swift\ narrow field telescopes.  We use the combined BAT and XRT
data to determine whether an extrapolation of the early gamma-ray
emission detected by the BAT joins smoothly to that seen by the XRT,
whether or not there is always a rapid decline phase seen by either
instrument and to investigate the relative importance of emission from
the central engine and the afterglow.

Our GRB sample is summarized in section 2. The analyses of the BAT and
XRT data are presented in sections 3 and 4 and we explain how the data
from the two instruments were combined in section 5. The observed
temporal and spectral shapes are presented in section 6. In section 7 we
describe a light curve fitting technique which we use to study the
various components contributing to the early X-ray emission. The
discussion and conclusions are given in sections 8 and 9 respectively.

\section{GRB sample}

Our initial GRB list of potential bursts comprised those detected by
\swift\ prior to 2005 October 1 for which \swift\ slewed to point its
narrow-field instruments within 10 minutes of the burst trigger
time. In the majority of cases the XRT observations started within 2
minutes.  Of the 45 such GRBs, we excluded 5 from our detailed
analysis: GRB050117A (T$_{90} = 167$s; Hill et al. 2006), GRB050509B
(T$_{90} = 0.04$s; Gehrels et al. 2005), GRB050815 (T$_{90} = 2.6$s),
GRB050906 (T$_{90} = 0.02$s) and GRB050925 (T$_{90} = 0.01$s). Four of
these are short and faint, and as a result insufficient X-ray photons
were obtained in the first hour with which to accurately constrain the
early X-ray spectral and temporal indices. Only one of the excluded
GRBs is a long burst (GRB050117A). Is this case the XRT obtained few
data due to it being observed initially whilst \swift\ was in the
South Atlantic Anomaly. Allowing for these problems, these bursts are
consistent with the sample studied here.

The 40 remaining GRBs form our sample and are listed in
Table~\ref{table1}.  Following convention, the bursts are named as
GRB-year-month-day with a following letter (A, B, C ...)  if multiple
bursts were detected on that day.  We adopt the usual T$_{90}$
convention for long and short bursts, but note that the assignment
does depend on the detector sensitivity and bandpass. We use the
15--150 keV band. We include 2 bursts --- GRB050724 and GRB050813 ---
which we classify as short bursts to see how they compare.  GRB050724
formally has a T$_{90}$ $>2$s in the BAT, but this is due to a long,
fairly soft X-ray tail of emission (Barthelmy et al. 2005b). It would
have appeared to be a short burst with T$_{90}
\sim 0.4$s in the BATSE/CGRO instrument (Fishman et al. 1994).

\section{BAT analysis}

All of the GRBs discussed in this paper were detected by the \swift\
Burst Alert Telescope. Once triggered, the BAT determines if there is
a new point source, and the \swift\ figure-of-merit processor then
computes whether this source can be observed immediately (i.e., is
unconstrained by the Earth, Moon or Sun).  If so, the satellite is
commanded to slew to put the target in the field of view of the
narrow-field instruments. The BAT continues to observe during the slew,
providing an uninterrupted light curve.

The BAT data for the GRB sample were processed using the standard BAT
analysis software (\swift\ software v. 2.0) as described in the BAT
Ground Analysis Software Manual (Krimm, Parsons \& Markwardt 2004) and
then light curves and spectra were extracted over 15--150 keV,
correcting the response matrix during slews\footnote{
{http:$//$swift.gsfc.nasa.gov$/$docs$/$swift$/$analysis$/$}}. 
The derived values of T$_{90}$ and T$_{50}$ are given in 
Table~\ref{table1},
along with the mean fluxes and spectral indices derived from fitting
single power laws ($N_{\rm ph}(E) \propto E^{-\Gamma_b}$) over the
period corresponding to T$_{90}$. Thus, the spectral properties are
averages over T$_{90}$. Throughout this paper all quoted errors on fit
parameters correspond to 90\% confidence for a single parameter
(i.e. $\Delta \chi^2 = 2.706$). {\tt XSPEC v11.3} (Arnaud 1996) was
used to fit the BAT and XRT spectra.

A single power law provides a statistically acceptable fit to the BAT
data in most cases. The four GRBs for which a cut-off power law model
($N(E) \propto E^{-\Gamma_{bc}} \exp({-E/E_{\rm cut}})$) provides a
significantly better fit (at $>99$\% confidence ) with a well
determined $E_{\rm cut}$ are noted in Table~\ref{table2}, which gives
the low energy spectral indices ($\beta_{bc} = \Gamma_{bc} - 1$) and
the e-folding energies of the exponential cut off ($E_{\rm cut})$. 
The T$_{90}$ and $\beta_b$ distributions are not significantly different
from those found for previous GRB samples.

\section{XRT analysis}

The XRT observations used here began at the times given in column 2 of
Table~\ref{table3}.  These times are relative to the GRB trigger time
determined by the BAT.  The XRT observations incorporate data taken
using the various operating modes of the instrument (see Hill et
al. 2004) and were corrected for pile-up where appropriate using the
method described in Nousek et al. (2006). The bulk of the XRT data
presented here were obtained using Windowed Timing (WT) or Photon
Counting (PC) modes (event grades 0--2 and 0--12 respectively).  The
XRT data were processed using {\tt xrtpipeline v0.8.8} into filtered
event lists which were then used to extract spectra and light curves
for the 0.3--10 keV energy range.

The X-ray light curves are generally complex, often with ``flares''
and frequent changes in temporal slope which can occur over short time
intervals, particularly at early times. As an initial simple
parameterisation, we have fitted the XRT light curves obtained over
the first few orbits with a broken power law model with flux $\propto
(t - t_0)^{-\alpha_1}$ before some break time, $t_{\rm break}$, and
$\propto (t - t_0)^{-\alpha_2}$ after the break.  This function
quantifies the early decay rate and any late emission hump. The
temporal decay slopes and the break times are given in Table~\ref{table3} and
were determined using the BAT trigger time as $t_0$. This time
corresponds to the start of the first foreground interval for which
the BAT was able to locate a point source. This may not coincide with
the initial rise in count rate.

For GRB50319 automatic triggering was disabled during the start of the
burst as \swift\ was slewing, and the GRB actually started some 135s
before the BAT trigger (Cusumano et al. 2006). For that GRB we include
all data following the onset of the burst determined from the pre- and
post-BAT-trigger light curve using the $t_0$ from Cusumano et
al. (2006).

We have visually judged when to exclude flares and short-duration
changes of slope to provide the early XRT decay index ($\alpha_1$).
We stress, however, that such a representation is indicative of,
rather than completely representative of, the complex X-ray light
curves. We return to this issue in section 7 where we adopt a more
automatic approach to parameterise the light curve and to select
$t_0$. Unlike Nousek et al. (2006), we do not distinguish in Table~\ref{table3}
between those bursts that decay rapidly or slowly in the earliest XRT
observations.  That is, we do not group bursts into decay phases based
on assuming an early steep decay phase is always visible --- we show
later that this is not always the case.  Rather the values of
$\alpha_1$ and $\alpha_2$ given in Table~\ref{table3} correspond to the decay
rates either side of the first clearly observed temporal break in the
XRT light curve.  In some bursts no break in X-ray temporal slope is
clearly detected using the XRT data alone, in which case only
$\alpha_1$ is tabulated.  The mean value of $\alpha_1$ is 2.45 with a
standard deviation of 1.55.

X-ray spectra were obtained from the early data and fitted with a
single power law model, allowing for both Galactic (Dickey \& Lockman
1990) and intrinsic absorption using the {\tt wabs} model in {\tt
XSPEC}. The default cross sections and abundances were used (Morrison
\& McCammon 1983; Andres \& Ebihara 1982).  The derived early XRT
spectral indices ($\beta_x$) and absorbing columns are given in
Table~\ref{table3}. In general little evidence is found for spectral
evolution across early temporal breaks (see also Nousek et al. 2006)
but where evolution is seen the XRT spectrum tends to get harder.  Aside
from GRB 050525A where photo-diode mode data were used, the data used
for the XRT spectra were obtained from the first orbit of WT data
(pre-break for GRB 050219A and pre-flare for GRB 050502B) or the first
orbit of PC mode data (GRBs 050126, 050128, 050315, 050319
(pre-break), 050401, 050406 (pre-flare), 050412, 050416A (pre-break),
050607, 050813, 050826, 050908 and 050916).

A majority of the GRBs show evidence for excess absorption above the
Galactic column at the $>99$\% confidence level. The excesses given in
Table~\ref{table3} have been converted to rest-frame absorbing columns
for those GRBs for which a redshift is known. For the bursts with
redshifts, the intrinsic column ranges from $2 \times 10^{21}$
cm$^{-2}$ up to $3.5
\times 10^{22}$ cm$^{-2}$ consistent with the idea that these GRBs
occur in a molecular cloud environment (Reichart \& Price 2002;
Campana et al. 2006).

\section{Combining the BAT and XRT light curves}

As the BAT and XRT data cover different energy bands and seldom
overlap in time, to compare them the data for one instrument must be
extrapolated into the bandpass of the other. We have chosen to use the
0.3--10 keV bandpass to determine fluxes as the bulk of the temporal
data, particularly during the decline, were obtained using the XRT.

\begin{figure}
\figurenum{1}
\centerline{
\includegraphics[scale=0.5]{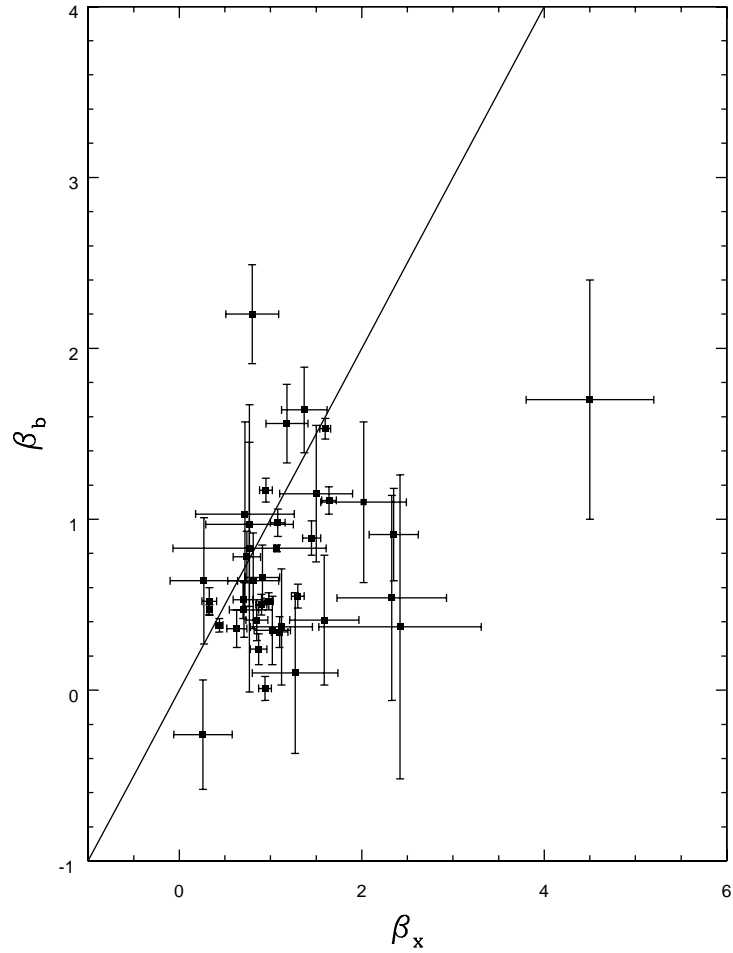}
}
\caption{
The relationship between the BAT ($\beta_b$) and early XRT ($\beta_x$)
spectral indices.  The solid line shows equality. This plot
illustrates that the early X-ray spectrum observed by the XRT tends to
be softer than the prompt gamma-ray emission observed by the BAT.  }
\label{figure1} \end{figure}

In Fig.~\ref{figure1} we plot of the distribution of $\beta_b$ and
$\beta_x$. The solid line shows equality of spectral index.  There is
a trend such that the early XRT data are fitted by a systematically softer
power law than the BAT data (i.e. $\beta_x > \beta_b$). The mean
values of $\beta_b$ (using single power law fits) and $\beta_x$ are
0.74 and 1.16 with standard deviations of 0.50 and 0.76 respectively.

\begin{figure}
\figurenum{2}
\centerline{
\includegraphics[]{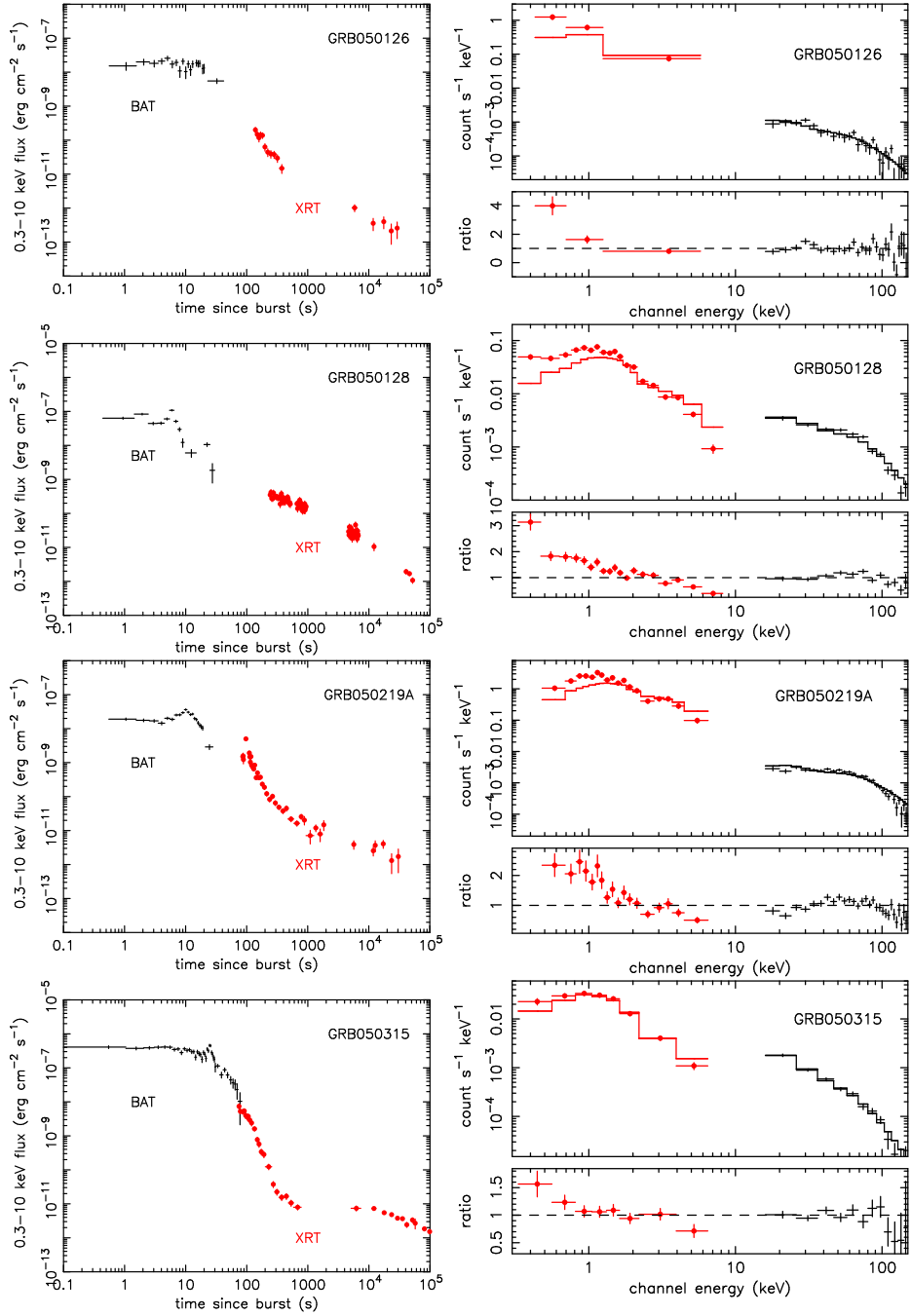}
}
\caption{
Left-hand panels show the combined BAT+XRT unabsorbed 0.3--10 keV flux
light curves plotted out to $10^5$s.  Right-hand panels show the
spectra relative to the power law derived from fitting the BAT data.
These plots were constructed as described in the text.  }
\label{figure2} \end{figure}

The spectral trend is in the same sense as the known tendency for GRB
prompt gamma-ray emission to become softer at later times (Ford et
al. 1995).  We have used this tendency when combining the BAT and XRT
data to form the unabsorbed, 0.3--10 keV flux light curves shown in
Fig.~\ref{figure2}.  These light curves were constructed by: (a)
converting the XRT count rates into unabsorbed fluxes using the power
law model parameters as given in Table~\ref{table3}; and (b)
converting the BAT 15--150 keV count rates into unabsorbed 0.3--10 keV
fluxes by extrapolating the BAT data to the XRT band using a power law
model with a spectral index which is the mean of the XRT and best-fit
BAT spectral indices.  This method was used for all bursts except
GRB050714B for which the early XRT spectral index is exceptionally
soft and appears to evolve rapidly (Levan et al. 2006).  For
GRB050714B we have used the $\beta_b$ from Table~\ref{table1} to
convert BAT counts to flux. It should be noted that in
Fig.~\ref{figure2} the time axis is limited at $10^5$s to emphasise
the early part of the light curve. For some bursts observations
continued beyond that time and all data were included when performing
the temporal fits given in Table~\ref{table3} and when analysing the
data further in section 6.

The observed spectra for the BAT and XRT and the ratio of the
spectra to a power law are also shown in Fig.~\ref{figure2}. In these
plots the relative normalisations of the XRT and BAT data were allowed
to be free parameters while the spectral index was frozen at the
single power law value obtained for the BAT. The residuals illustrate
the generally softer X-ray spectrum seen by the XRT compared to the
BAT.

\section{The observed early temporal and spectral shape}

The observed GRB properties for the \swift\ sample in terms of
spectral hardness and duration indicate that it is broadly
representative of the GRB population observed by BATSE (Kouveliotou et
al. 1993; Berger et al. 2005). The difference is that \swift\
provides X-ray data over a much longer time interval than previous
missions and can detect GRBs with lower mean fluxes.  Combining the BAT
and XRT data also allows for a view of the prompt phase of GRB emission with
little or no temporal gap.

Although GRB light curves can have considerable structure superimposed
on the overall decline (see below) following the initial burst, the
data shown in Fig.~\ref{figure2} strongly suggest that there is no
discontinuity between the emission seen by the BAT and the later
emission seen by the XRT.  In every case where the burst was long
enough for the XRT to start observing before the end of T$_{90}$ the
BAT and XRT light curves join smoothly. For those GRBs where there is
a temporal gap of only a few tens of seconds the light curves can be
smoothly extrapolated to join.  For those GRBs with a short T$_{90}$
the extrapolation is naturally over a longer time, but even for these
cases the BAT and XRT data appear to join up.

The only cases for which there is a temporal gap and where the
extrapolated BAT light curve may not agree with the XRT are GRB050219A
and possibly GRB050525A. For GRB050219A, allowing for the spectral
evolution observed by the BAT does not resolve this problem
(Tagliaferri et al. 2005; Goad et al. 2005). The apparent
discontinuity could, however, be due to an X-ray flare in this burst
around 90s after the trigger.  As noted above, the presence of late
X-ray flares in GRBs is now known to be common. For GRB050525A, which
like GRB050219A is best-fitted by a cut-off power law in the BAT, some
spectral evolution or an early break in the light
curve similar to that in GRB050713A could be responsible.

For the 5 bursts excluded from detailed analysis, no 
counts were detected with the XRT for GRB050906 and GRB050925. The
combined BAT+XRT X-ray light curve for GRB050815 suggests it declined
rapidly before 100s and then decayed more slowly until a few
thousand seconds, similar to the X-ray decay seen in GRB050509B (Gehrels et
al. 2005).  GRB050117A, which was observed while \swift\ was in the
South Atlantic Anomaly, appears to display a rapid decline after the
prompt phase followed by a shallow decline until 10ks (Hill et
al. 2006), and hence is consistent with the behaviour of the rest of
the long bursts in the sample.

As GRBs are, by their very nature, transient, all bursts show a
decline following the prompt phase. The rate of decline, and how long
it lasts, varies. This is indicated by the large standard deviation in
$\alpha_1$ noted above. Visual inspection of the light curves show
that some two-thirds of the GRBs have X-ray light curves that show an
early more rapid phase of decline which then breaks to a late emission
hump. This temporal break is observed to occur over a wide range of
times but is usually within the first hour, and is clearly not a jet
break. The remaining GRBs seem to have a continuous decline. Those
which do not seem to flatten have only a single ($\alpha_1$) decay
index given in Table~\ref{table3}. The mean value of the X-ray
spectral index is not different between those GRBs with or without a
rapid decay phase.

Around half of the GRBs appear to show late ($t >$ T$_{90}$) X-ray
flares. These include GRB050406, 050502B, 050607, 050713A,
050714B, 050716, 050724, 050801, 050813, 050820A, 050822, 050904,
050908, 050916 and 050922B. A few others (e.g. GRB050219A, 050319,
050802, 050915A) may have flares at the start of the XRT
observation. Most of these flares contain the equivalent of 10\% or
less of the prompt fluence, but in a few cases have $> 50$\%
(e.g. GRB050502B (Burrows et al. 2005b) and GRB050820A (Osborne et
al. 2006)). There is no significant difference between the rate of
flares for those GRBs with or without a very steep decline X-ray
phase.

\begin{figure}
\figurenum{3}
\centerline{
\includegraphics[scale=0.5]{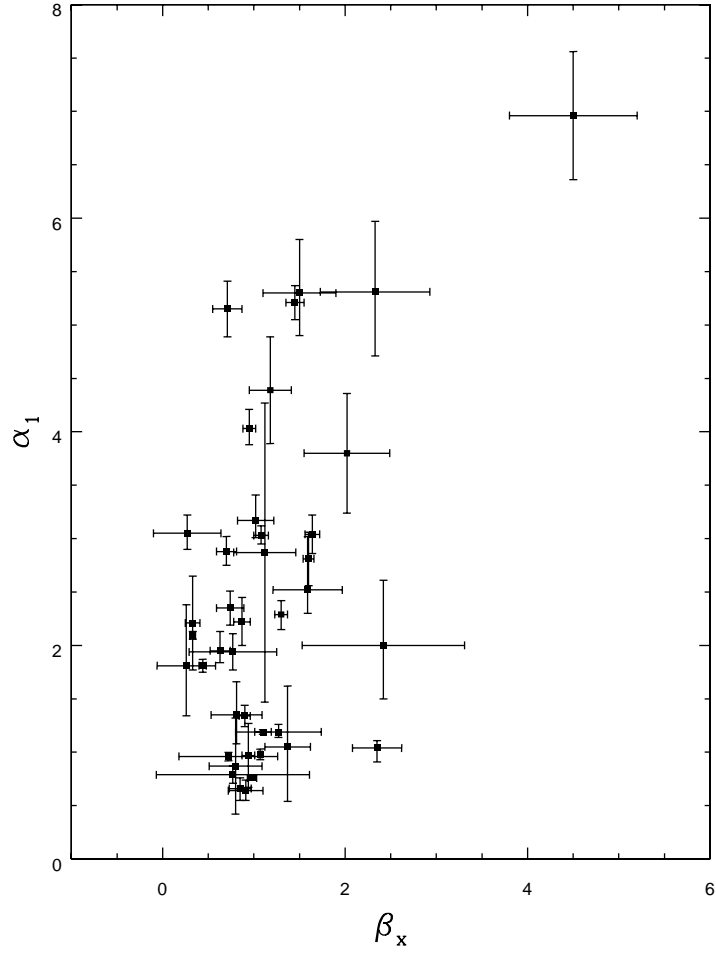}
}
\caption{
The relationship between the observed X-ray temporal ($\alpha_1$) and
spectral ($\beta_x$) indices. The outlier at top-right is GRB050714B.}
\label{figure3} \end{figure}

\begin{figure}
\figurenum{4}
\centerline{
\includegraphics[scale=0.5]{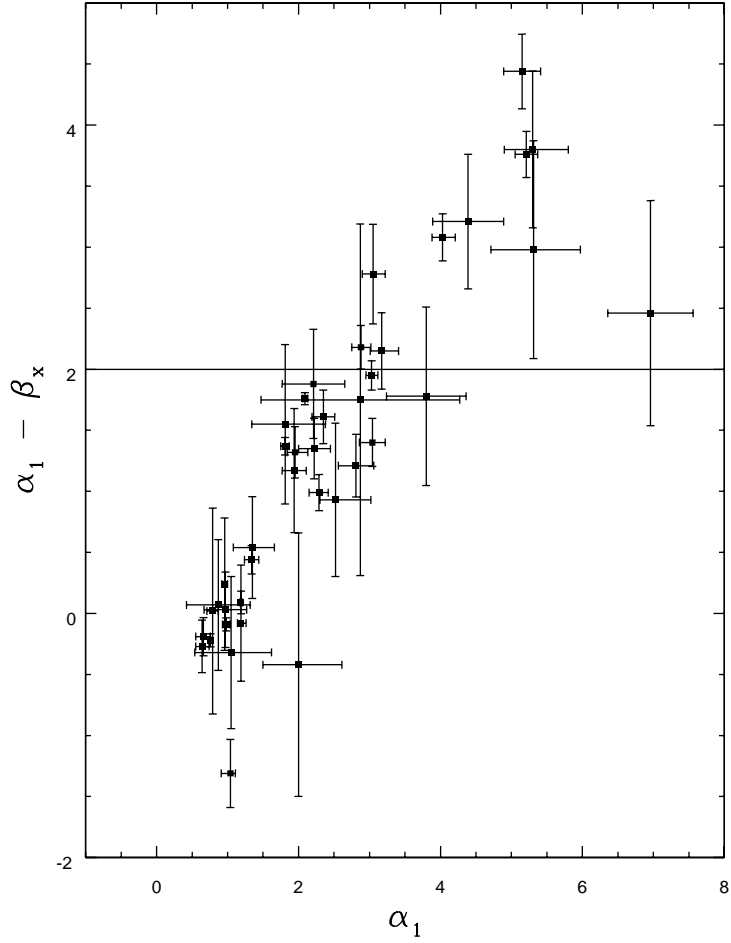}
}
\caption{
The difference between the observed X-ray temporal ($\alpha_1$) and
spectral ($\beta_x$) indices, derived from the early XRT data, as a
function of temporal index.  The horizontal line shows $\alpha_1 -
\beta_x = 2$ as predicted by the high latitude emission model 
(Kumar \& Panaitescu 2000). }
\label{figure4} \end{figure}

Neither $\alpha_1$ nor $\beta_x$ are significantly correlated
with T$_{90}$. There is a weak correlation between $\alpha_1$ and
$\beta_x$ (Fig.~\ref{figure3}; Spearman's rank correlation coefficient
$r = 0.30$, significant at 95\%), although this depends on including
GRB050714B. In contrast, over a wide range in $\alpha_1$, there is a
very strong correlation between ($\alpha_1 -
\beta_x$) and $\alpha_1$ (Fig.~\ref{figure4}, $r=0.89$, significant at $\gg
99.9$\%).  The sense of the correlation is such that those GRBs with
shallower temporal decays have a smaller difference between their
temporal and spectral indices. This correlation is consistent with no
strong dependence of $\beta_x$ on $\alpha_1$.

It has been suggested (e.g. Nousek et al. 2006; Zhang et al. 2006) that the
steeply declining early X-ray emission seen in early XRT observations
can be interpreted in terms of ``high latitude'' emission (Kumar \&
Panaitescu 2000). In this model if the physical process producing the
X-ray emission (such as internal shock activity) stops, the observed
emission will continue for a short time as the observer will continue
to see emission from those parts of the jet which are off the
immediate line of sight. Thus emission at angles $\theta$ from the
line of sight which are in excess of $\theta = \Gamma_{jet}^{-1}$ will
start to dominate the observed emission, where $\Gamma_{jet}$ is the
jet Lorentz factor.  For a uniform surface brightness jet, the energy
in some bandpass will fall as $t^{-\beta - 2}$ where the spectrum is
$\propto \nu^{-\beta}$ (Kumar \& Panaitescu 2000).  In terms of
spectral index, this model predicts a relation such that $\alpha -
\beta =2$ for the early, rapidly declining part of the temporal decay.
It is possible to get a shallower decay if viewing a structured-jet
off-axis (Dyks, Zhang \& Fan 2006) although the
general trend is similar to the standard high-latitude model.

From Fig.~\ref{figure4}, for the entire sample $\alpha_1$ appears to be largely
independent of $\beta_x$ or at least is not simply related. This is
inconsistent with the concept that the early X-ray emission is usually
dominated by high latitude emission in a uniform jet. An alternative
explanation for the early X-ray emission GRBs is afterglow emission,
but this also predicts a specific relation between the spectral and
temporal indices, as discussed below. Overall, the data shown in
Fig.~\ref{figure4} suggest that no single model will explain all of the bursts.

To try and force a fit with a particular model we could adjust the
start time of the time series, $t_0$, to some point in the prompt
light curve such that the relation between $\alpha$ and $\beta$ fits a
particular model and is not simply defined by the BAT trigger time..
For example, moving forwards to a late flare, if one is observed, or
earlier to a visually estimated start time. Rather than adopt such a
subjective and model dependent approach, we have developed a method to
automatically align the light curves. In sections 7 and 8 we use this
technique to investigate the early X-ray light curve in GRBs and hence
determine the contribution of likely emission components.

\section{The global GRB X-ray decay curve}

In order to compare light curves for different GRBs and to study the
different phases of the X-ray emission, we have developed a procedure
to fit light curves. This procedure, described below, attempts to
provide a best fit to the ``global light curve'' for all our GRBs,
assuming there is a generic pattern to their behaviour. It uses
information from the combined BAT+XRT flux light curve for each burst and
can adjust the start of the time series (i.e. move $t_0$) and the temporal
scale of the decay, while also allowing for deviations from the global
decay (termed flares and humps below) to derive a standard
set of parameters for each burst.

An average X-ray decay curve expressed by log(time) as a function of
log(flux), $\tau(F)$, and log(flux) as a function of log(time),
$F(\tau)$, was derived by taking the sum of scaled versions of each of
the individual light curves, $f_{i}(t_{i})$, where $t_{i}$ is
approximately the time since the largest/latest peak in the BAT light
curve.  The data points were transformed to normalised log(flux),
$F_{i}=\log _{10}(f_{i}/f_{d})$, and log(time) delay values,
$\tau_{i}=\alpha_{d} \log_{10}(t_{i}-t_{d})-\tau_{d}$.  Four decay
parameters (suffix {\it d}) specify the transformation for each GRB:
$f_{d}$, the mean prompt flux; $t_{d}$, the start of the decay;
$\tau_{d}$, a time scaling; and $\alpha_{d}$, a stretching or
compression of time.  The flux scaling is a simple linear shift in the
log-scale and does not involve any stretching or compression of one
GRB with respect to another.  Under the transformation all the light
curves conform to an approximately universal behaviour with an initial
exponential decline $\propto \exp(-t/t_{c})$ followed by a power law
decay $\propto t^{-\alpha_{0}}$. 

The transition between the two decay phases occurs when the
exponential and power law functions and their first derivatives are
equal, and is given for the average decay curve by
$t_{0}=t_{c}\alpha_{0}$ ($\tau_0 = 1.7$). Adopting this transition,
for each GRB we define the division between the prompt and power law
decay phases to be $\tau_{0}$, corresponding to a prompt time
T$_{p}=10^{(\tau_{0}+\tau_{d})/\alpha_{d}}$ seconds.  This definition
of T$_{p}$ provides us with an alternative estimate of the duration of
each burst which depends on the physical shape of the light curve,
takes into account the data from both the BAT and the XRT and is not
bound by the sensitivity of either instrument. It does depend on the
chosen energy band and method used to extrapolate. Depending on the
particular characteristics of the burst, T$_{p}$ can be similar to,
greater than or smaller than T$_{90}$.

The best fit $f_{d}$, $t_{d}$, $\alpha_{d}$ and $\tau_{d}$ for each
GRB were found using a least squares iteration procedure. Initial
values for these parameters were chosen by visual inspection of the
light curves and a first guess for the average decay curve,
$F_{av}(\tau_{av})$, was set up using nominal values for $t_{c}$ and
$\alpha_{0}$.  The zero time $t_{d}$ was initially set to the position
of the largest peak for each burst.  The flux was normalised by
calculating $f_{d}$ which minimized the square difference between the
average flux curve and each data point,
$\sigma_{f}=\sum(F_{i}-F_{av}(\tau_{i}))^{2}$, summing over all data
points in the prompt phase $\tau<\tau_{0}$.
In order to perform the least squares fitting in the time domain, each
light curve was binned as temporal delay values over a set of bins in
$F$ (rather than averaging the flux values over time bins) and the
average temporal delay for each flux bin was calculated.
The temporal parameters $\tau_{d}$,
$\alpha_{d}$ and $t_{d}$ were updated minimizing
$\sigma_{\tau}=\sum(\tau_{av}(F_{i})-\tau_{i})^{2}$ summing over all
the measured values at different flux levels for
$\tau_{0}<\tau<\tau_{h}$ where $\tau_{h}=3.5$. The upper limit
$\tau_{h}$ is set to take into account the late emission hump. Bright
flares in the decay phase were excluded during the minimization
procedure. 

\begin{figure}
\figurenum{5}
\centerline{ 
\includegraphics[scale=0.7,angle=-90]{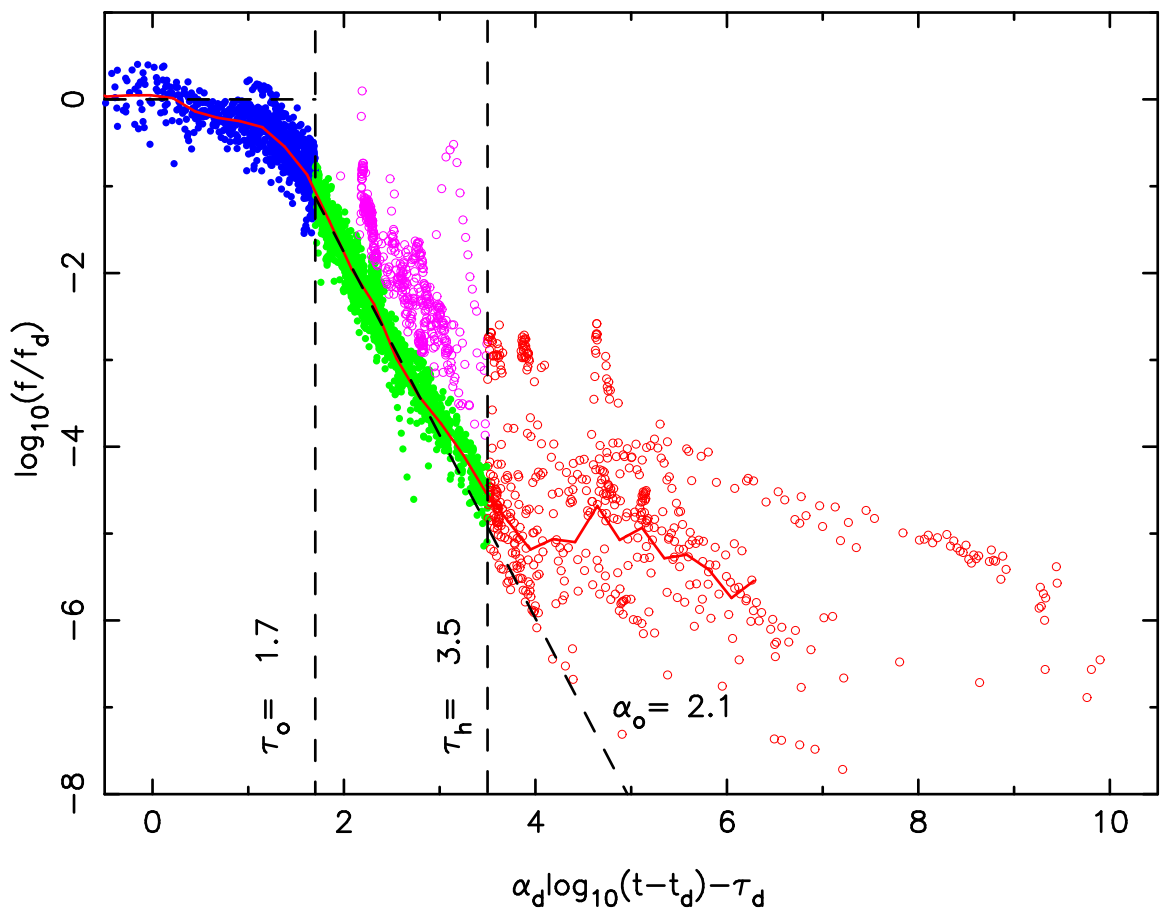}
}
\caption{
Composite decay curve derived from 40 GRBs. See text for details. The
filled (blue) circles at $\tau < 1.7$ are within time T$_p$.  The
average produced by the least squares procedure is shown as the solid
(red) line.  The dashed power law slope indicates the best fit to the
initial decay in the average curve derived using the filled (green)
circles. The open circles indicate flux measurements which were
excluded from the power law fitting as flares (pink) or because
$\tau>3.5$ (red).  }
\label{figure5} \end{figure}

\begin{figure}
\figurenum{6}
\centerline{
\includegraphics[scale=0.7,angle=-90]{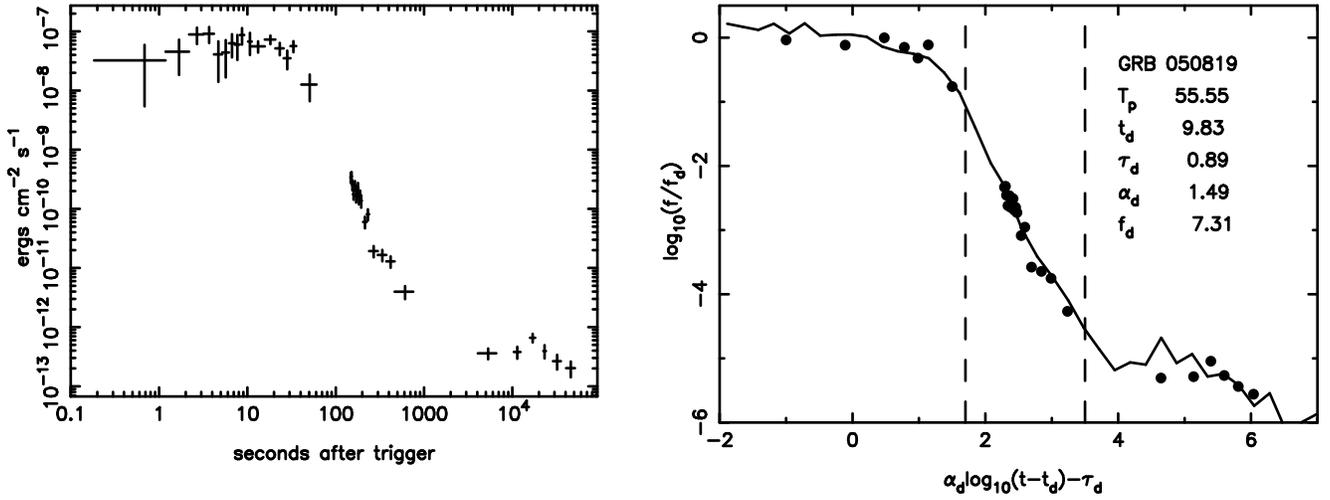}
}
\caption{
Left-hand panel: The unabsorbed 0.3--10 keV flux light curve for GRB 050819. 
Right-hand panel: transformed light curve with best fit parameters.
Vertical lines and solid curve as in Fig.~\ref{figure5}.}
\label{figure6} \end{figure}

At the end of each iteration the $\alpha_{d}$ values were scaled so
that the mean over all the GRBs was unity and the $\tau_{d}$ values
were offset so that the mean was zero. This ensures that $\tau$ has
units of $\log _{10}(seconds)$. Using the updated parameter values a
new best guess at the average decay curve was calculated summing over
all 40 GRBs. The iteration was stopped when the changes in the
parameters were sufficiently small.  The average decay curve derived
from 40 GRBs and all the measured flux values from the BAT and XRT
(4569 measurements) are shown in Fig.~\ref{figure5}. The values of
T$_{p}$ are given in Table~\ref{table4}. An example of the
transformation for one burst (GRB 050819) is shown in
Fig.~\ref{figure6}. This figure shows that $t_d$ is moving the zero
time (hence some of the early prompt points are before the start time in
transformed space).  Then $\tau_d$ and $\alpha_d$ rescale the units of
logarithmic time.

The average decay curve relaxes into a power law with a decay index
$\alpha_{0}=2.1$ found by linear regression on the average decay curve
for $\tau_{0}<\tau<3.0$. This power law fit is shown as a dashed line
in Fig.~\ref{figure5}. 
The fitting procedure results in those GRBs which follow a
fairly continuous decay lying close to the power law. At $\tau \sim
3$ the average decay curve starts to rise above the power law decay in
the majority of bursts. This is the start of the late emission hump,
which has a large observed variety in strength which we parameterize
below. The range in hump strength results in the somewhat jagged
appearance of the average decay curve for $\tau > 3.5$.

GRB050730 falls below the average decay curve at $\tau \sim 2.5$. This
object has the largest value of T$_{p}$ (373 seconds) and a decay
which gets much steeper at $\tau\approx2.5$. The last few data points
for this source can be seen below the bulk of the data on
Fig.~\ref{figure5}.  For GRB050319 the fit procedure prefers a fit
using the first (larger) peak in the BAT light curve as the burst and
treats the second, later peak as a flare. If we force the procedure to
adopt a $t_0$ just before the second lower-flux peak we derive a very
steep early decay index ($\sim 5$) but get a significantly worse fit
to the rest of the light curve.

\begin{figure}
\figurenum{7}
\centerline{ 
\includegraphics[scale=0.7,angle=-90]{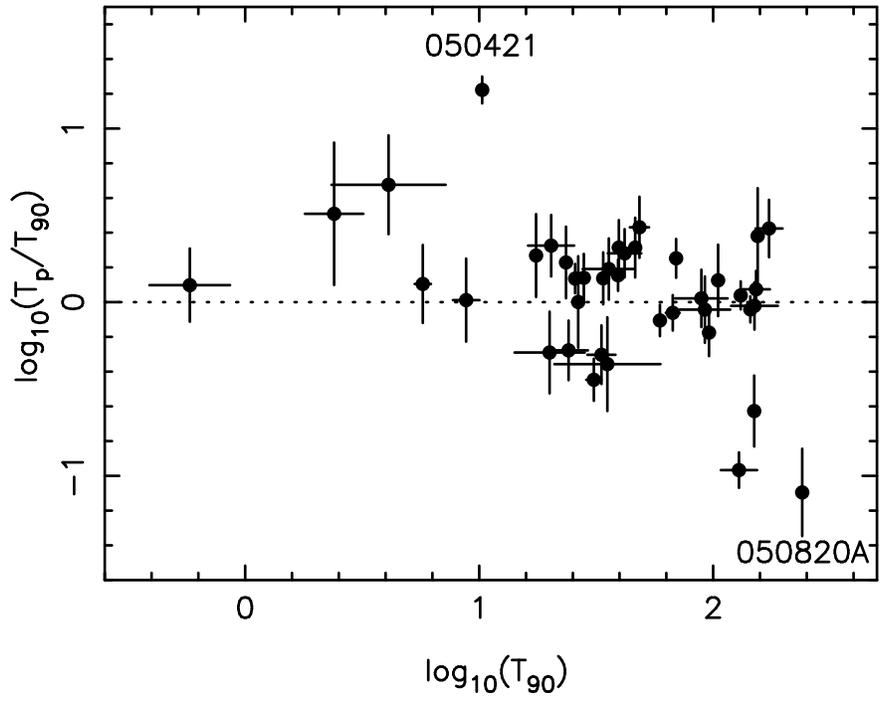}
}
\caption{The
prompt time T$_{p}$ compared with T$_{90}$.
}
\label{figure7} \end{figure}

The distribution of $\log_{10}($T$_{p}/$T$_{90}$) as a function of
$\log_{10}($T$_{90}$) is shown in Fig.~\ref{figure7}.  For almost all bursts
T$_{p}$ is comparable to or somewhat larger than T$_{90}$ as
expected. For GRB050421 it is a factor of 17 larger.  For this burst,
if the BAT data beyond the T$_{90}$ period are binned up a long tail
extending to the start of the XRT observations can be seen (Godet et
al. 2006). For GRB050820A T$_{p}$ is a factor of 12 smaller than
T$_{90}$.  There is a very bright second burst/flare seen from this
object which was excluded from the average curve fitting but which is
included when calculating the BAT T$_{90}$ estimate. This second event
is brighter than the first (Osborne et al. 2006).

\begin{figure}
\figurenum{8}
\centerline{ 
\includegraphics[scale=0.7,angle=-90]{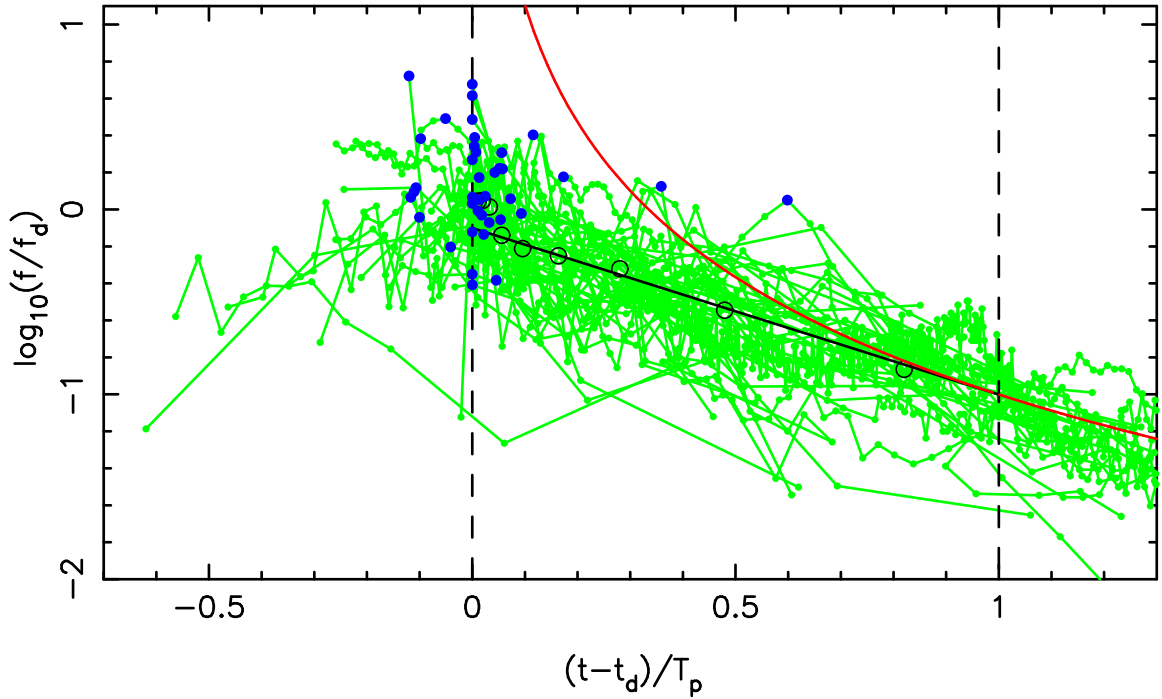}
}
\caption{The
distribution of peak flux times (filled (blue) circles) about time
zero produced by the least-squares fitting. The connected open circles
show the average decay curve relative to the individual (green) light
curves plotted in linear time relative to T$_p$. The curved (red)
solid line shows the backwards extrapolation of the average power law
which fits times above T$_p$. The vertical dashed lines correspond to
0 and T$_{p}$.}
\label{figure8} \end{figure}

The early light curves are plotted in linear time relative to T$_p$,
$(t-t_{d})/{\rm T}_{p}$ in Fig.~\ref{figure8}.  The peaks (global
maximum flux measurements for each burst) are shown as filled circles.
The notable exception is GRB050422 (Beardmore et al. 2006) for which
the peak occurs at $(t-t_{d})/{\rm T}_{p} \approx 0.6$. For this burst
a large late flare is seen in the BAT followed by a rapid decay. In
this case the fitting procedure has chosen $t_{d}$ such that the peak
falls at the centre of the T$_{p}$ window. GRB050922B also has a
bright flare late in the BAT light curve giving a peak at
$(t-t_{d})/{\rm T}_{p} \approx 0.35$.  Because the peaks are clustered
around zero we can be confident that any temporal indices derived for
the subsequent decays relate to the appropriate maximum in the light
curve.  The line connecting the open circles is a linear fit to the
average decay curve. Since this is a linear-log plot this represents
an exponential decay from the initial peak value. The time constant
for the exponential decay is $t_c = 0.47{\rm T}_{p}$ which is
equivalent to $\tau=1.4$ on Fig.~\ref{figure5}. The curved solid line
is the extrapolated power law.

\begin{figure}
\figurenum{9}
\centerline{ 
\includegraphics[scale=0.7,angle=-90]{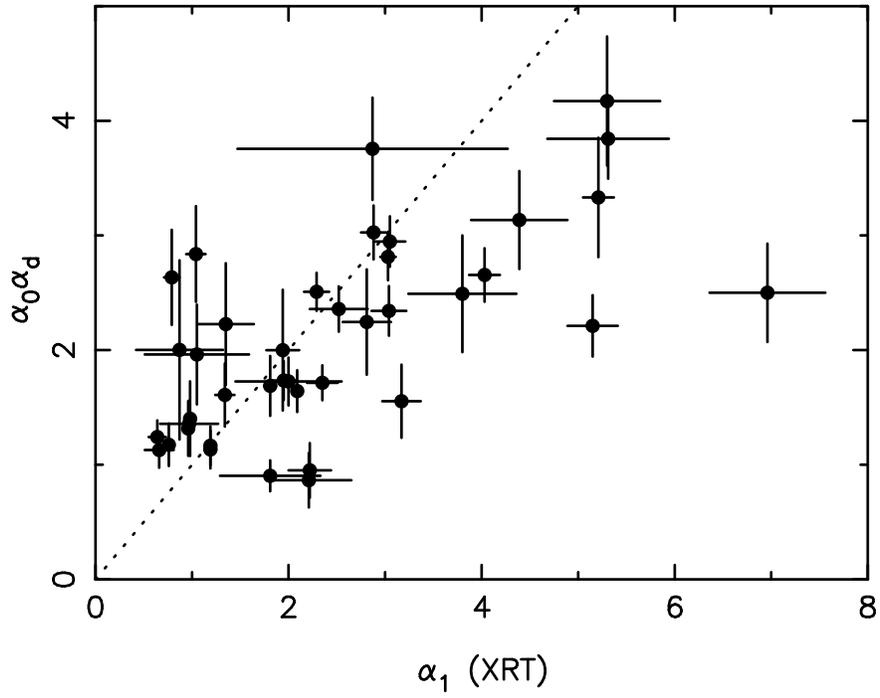}
}
\caption{
Correlation of $\alpha_{0}\alpha_{d}$ derived from the global light
curve fit with $\alpha_1$ estimated from the XRT light curve. The
dotted line shows $\alpha_{0}\alpha_{d} = \alpha_1$.}
\label{figure9} \end{figure}

The initial temporal decay index for individual GRBs can be calculated
by multiplying $\alpha_{0}$ by the best fit $\alpha_{d}$.  GRBs with
$\alpha_{d}>1$ have decays steeper than average and those with
$\alpha_{d}<1$ shallower.  In Fig.~\ref{figure9} we plot
$\alpha_0\alpha_d$ versus the $\alpha_{1}$ parameter derived from the
XRT light curves (from Table~\ref{table3}).  Using the least squares
fit to the average decay curve has reduced the spread of decay index
values and in general the $\alpha_{0}\alpha_{d}$ are somewhat smaller
than $\alpha_{1}$.  The $\alpha_{0}\alpha_{d}$ are based on all the
available data from both the BAT and XRT and are expected to be a more
robust estimate of the global average. In contrast $\alpha_{1}$
provides a snap-shot of a particular (often rather short) section in
the overall light-curve and were derived using the BAT trigger time.

Many of the GRBs have excess flux over and above the average power law
decay. In order to quantify this we have calculated the average
difference, $\sum(F_{i}-F_{pl}(\tau_{i}))/n_{r}$, between the measured
data and the average power law decay $F_{pl}(\tau)$ over the ranges
$\tau_{0}<\tau<\tau_{h}$, giving $\Delta_{F}$, and over $\tau_{h}<\tau<10$,
giving $\Delta_{H}$, summing over the $n_{r}$ data points which fall
within each range.  Thus $\Delta_{F}$ and $\Delta_{H}$ provide a
measure of the flaring activity in the power law decay phase and the
strength of the late emission hump. The mean differences are
calculated in $\log_{10}$(flux) and therefore represent $\log_{10}$ of
a multiplicative factor over and above (or below if negative) the
average power law decay curve. Two bursts, GRB050202B and GRB050916
have a flare at $\tau > \tau_{h}$ which is included in their
$\Delta_H$ calculation.

\begin{figure}
\figurenum{10}
\centerline{ 
\includegraphics[scale=0.7,angle=-90]{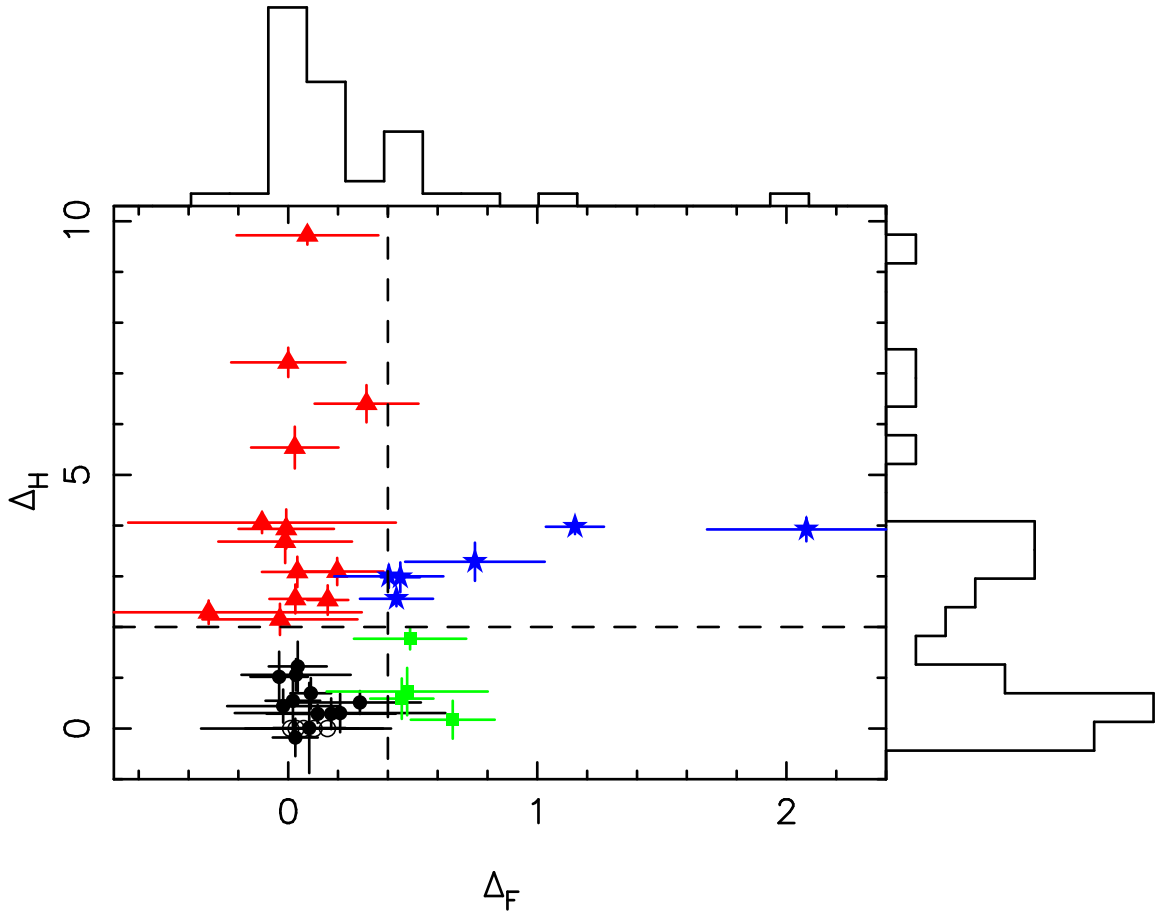}
}
\caption{
The distribution of $\Delta_F$ (measure of flaring activity) and $\Delta_H$
(measure of size of late hump).
The GRBs in each quadrant are shown as filled (black)
circles (no significant flares or hump), filled (green) squares
(flares but no hump), filled (blue) stars (flares and hump) and
filled (red) triangles (hump but no flares).
The open circles
denote GRBs for which there are no data in the light curve with $\tau > 3.5$.
}
\label{figure10} \end{figure}

The values of $\Delta_{H}$ plotted against $\Delta_{F}$ are shown in
Fig.~\ref{figure10}.  There is no correlation between these values.
The GRBs in each quadrant are shown as filled (black) circles (no
significant flares or hump), filled (green) squares (flares but no
hump), filled (blue) stars (flares and hump) and filled (red)
triangles (hump but no flares). The objects plotted as open circles
(with $\Delta_{H}=0$) are those for which there are no late data. In
most cases this is because the afterglow was too faint to detect in
the XRT and so a value of zero has been taken as a reasonable estimate
of $\Delta_{H}$ for these objects.

\begin{figure}
\figurenum{11}
\centerline{ 
\includegraphics[scale=0.7,angle=-90]{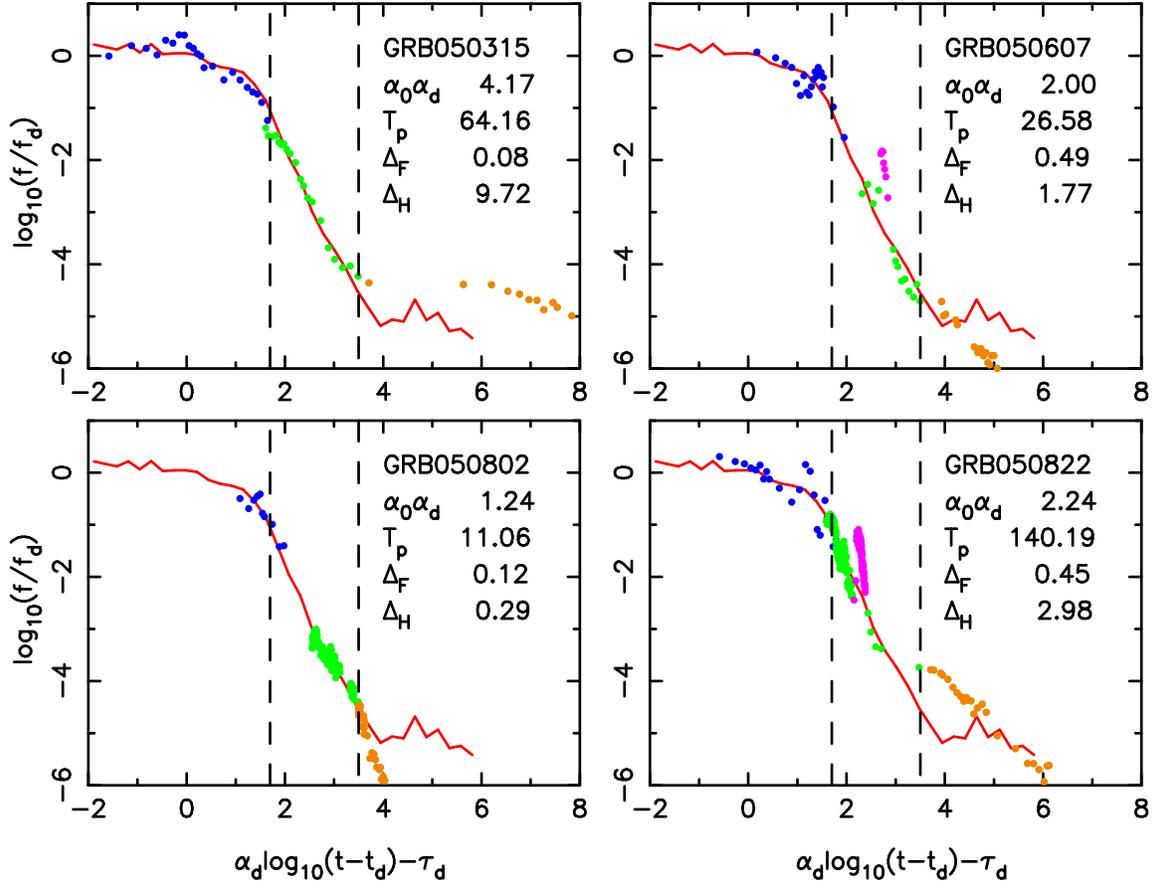}
}
\caption{
Examples of scaled GRB light curves fitted to the average decay curve.
Top-left panel: No flares with prominent late hump.
Top-right panel: Flares with weak late hump.
Bottom-left panel: No flares or late hump.
Bottom-right panel: Prominent flares and moderate late hump.
BAT data are filled blue circles, XRT data
are filled green circles. Flares and hump are filled purple
and orange circles respectively. Vertical dashed lines as in 
Fig.~\ref{figure5}.}
\label{figure11} \end{figure}

The values of $\alpha_{0}\alpha_{d}$, $\Delta_{F}$, $\Delta_{H}$,
T$_{p}$ and $\alpha_f$ (defined below) for the GRB sample are given in
Table~\ref{table4}.  Fig.~\ref{figure11} shows four examples of scaled
GRB light curves plotted with the average decay curve. These examples
illustrate GRBs with a range of early decay rates and flares and humps
of different strengths.

\section{Discussion}

The procedure which calculates the average X-ray decay curve generates
several parameters for each GRB: $f_{d}$, the mean prompt flux level;
T$_{p}$, the duration of the prompt emission;
$\alpha_{0}\alpha_{d}$, the temporal decay index of the initial decay;
$\Delta_{F}$, a measure of the level of flaring activity during the
initial power law decay; and $\Delta_{H}$, a measure of the size of the
late emission hump after the initial power law decay.

\begin{figure}
\figurenum{12}
\centerline{ 
\includegraphics[scale=0.7,angle=-90]{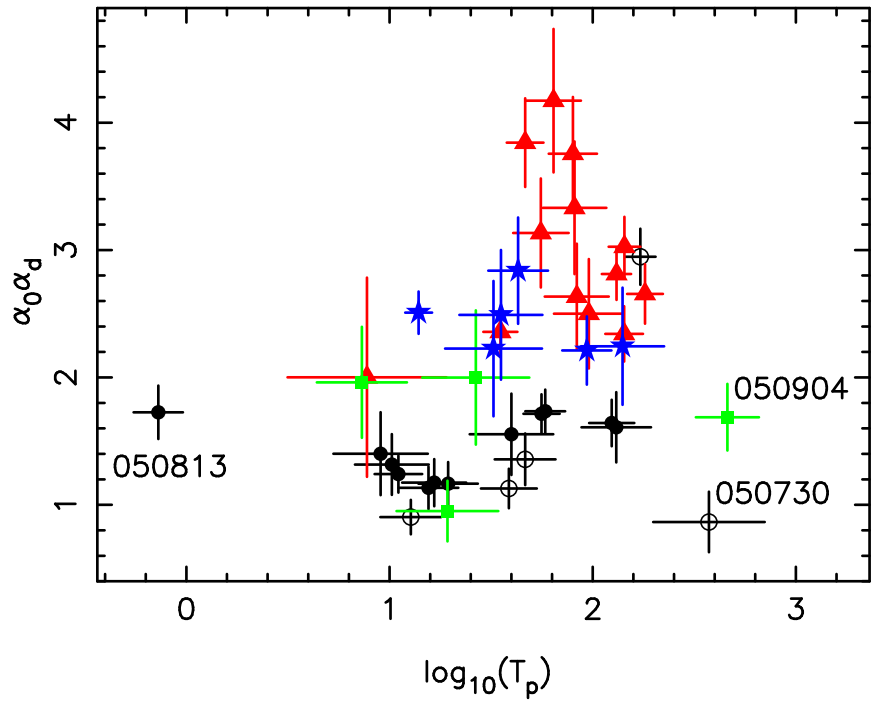}
}
\caption{
Distribution of $\alpha_{0}\alpha_{d}$
and $\log_{10}$(T$_p$). Symbols as in Fig.~\ref{figure10}.
}
\label{figure12} \end{figure}

The distributions of $\alpha_{0}\alpha_{d}$ and $\log_{10}({\rm
T}_{p})$ are shown in Fig.~\ref{figure12}. There is no correlation
between the duration of the burst and the initial decay
index. However, the bursts clearly cluster according to the relative
prominence of flares or humps.

As shown in Fig.~\ref{figure12}, the majority of the sample are
bunched in the range $4.5<{\rm T}_{p}<140$ seconds and
$0.9<\alpha_{0}\alpha_{d}<3.2$. GRB050730 (T$_{p}=157$,
$\alpha_{0}\alpha_{d}=0.85$) and GRB050904 (T$_{p}=282$,
$\alpha_{0}\alpha_{d}=1.39$), both high redshift bursts, are somewhat
isolated from the main population with relatively long decays and low
temporal decay indices.  GRB050813, a short burst, has the smallest
T$_p$.

\begin{figure}
\figurenum{13}
\centerline{ 
\includegraphics[scale=0.7,angle=-90]{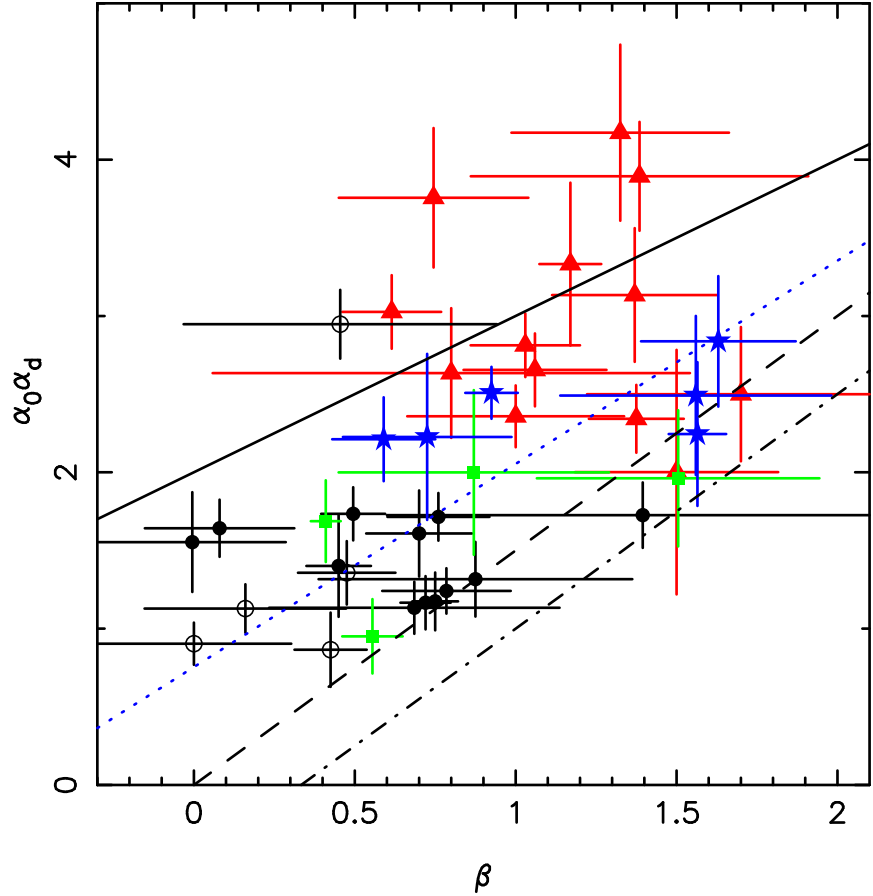}
}
\caption{
Correlation of $\alpha_{0}\alpha_{d}$ with $\beta$, where $\beta$ is
the average of the spectral indices from the BAT and XRT. Symbols as
in Fig.~\ref{figure10}.  The solid diagonal line shows the predictions of 
the high
latitude model. The dashed and dot-dashed diagonal lines show the two
afterglow models discussed in the text. The dotted diagonal line shows
a fit to those bursts below the high latitude line.  }
\label{figure13} \end{figure}

We can use the values of $\alpha_{0}\alpha_{d}$ to test the high
latitude and other emission models.  The correlation of
$\alpha_0\alpha_d$ with $\beta$, the average of the BAT and XRT
spectral indices (except for GRB050714B for which $\beta_{b}$ was
used), is shown in Fig.~\ref{figure13}. There is a correlation
(correlation coefficient 0.53, significant at $99$\%) between the
decay index and the spectral index for all the data plotted, but the
relation between $\alpha_0\alpha_d$ and $\beta$ does not match the
high latitude model prediction, shown by the solid line, and there is
clearly very significant scatter. This correlation, and other
significant correlations discussed in this paper, are summarized in
Table~\ref{table5}.

In principle, the relationship between the temporal decay index and
spectral index has two components such that
$\alpha=\alpha_{\nu}\beta+\alpha_{f}$. The coefficient $\alpha_{\nu}$
arises from the redshift of the peak of the spectral distribution of
the synchrotron emission as a function of time, while $\alpha_{f}$ arises
from the temporal decay in the peak flux value of the same spectral
distribution.  The solid line in Fig.~\ref{figure13} shows the
expected relationship for the high latitude model (Kumar \& Panaitescu
2000) with $\alpha_{\nu}=1$ and $\alpha_{f}=2$. The dashed line shows
the relationship expected from an afterglow expanding into a constant
density ISM observed at a frequency below the cooling break ($\nu_x <
\nu_c$) and before a jet break, with $\alpha_{\nu}=3/2$ and
$\alpha_{f}=0$ (Sari, Piran \& Narayan 1998). We will use this as our
afterglow model. If $\nu_x > \nu_c$ then $\alpha_{\nu}$ is unchanged
and $\alpha_f = -0.5$; this is plotted as a dot-dashed line on
Fig.~\ref{figure13}. All of the GRBs lie on or above these afterglow
lines. We adopt the ISM model as the light curves for most GRBs appear
consistent with the presence of a fairly constant density medium
(e.g. Panaitescu \& Kumar 2002, Yost et al. 2003, Blustin et
al. 2006). Using a wind model would not change the conclusions below.

The best fit correlation for all the bursts shown in Fig.~\ref{figure13} 
has $\alpha_{\nu}=1.8$ and $\alpha_{f}=0.53$, which is a poor fit to 
either model.  The best fit correlation (correlation coefficient 0.66, 
significant at $\gg 99.9$\%) for those bursts which lie below the high 
latitude line is shown as the dotted line.  For these objects 
$\alpha_{\nu}=1.3$, close to the average of the high latitude and 
afterglow models, but the intercept is $\alpha_{f}=0.75$, indicating that 
the peak flux is not decaying as fast as expected for the high latitude 
emission.

Of the 5 GRBs which lie significantly above the high latitude
prediction in Fig.~\ref{figure13}, 4 have the most significant late
humps. The other is GRB050421 for which we have no late data as it
quickly faded below detectability (Godet et al. 2006). Aside from
GRB050421, 3 out of 4 have $\alpha_{0}\alpha_{d} > 3$ which is
formally the maximum that the high latitude model allows (Kumar \&
Panaitescu 2000). One of these, GRB050422, has a large late flare in
the BAT which is placed at $(t-t_{d})/T_{p}\approx 0.6$ (see
Fig.~\ref{figure8}). If this peak is pushed nearer to the temporal
origin then the decay index becomes smaller ($\approx 3.2$, Beardmore
et al. 2006) but if we do this the overall fit to the average decay
profile is very poor. For the other two, GRB050315 and GRB050915A, the
most significant peak in the BAT light curve is close to the temporal
origin and the overall fit to the average profile is excellent. The
BAT light curve for GRB050315 is dominated by a single flare and
Vaughan et al. (2006a) give a similarly high decay index value
($\alpha_{1}\sim 5$). The large majority of GRBs lie below the high
latitude prediction. For these it is likely that we are seeing a
combination of high latitude prompt emission and conventional,
pre-jet-break afterglow.

The lines shown on Fig.~\ref{figure13} are just some of a family of
curves with the form $\alpha=\alpha_{\nu}\beta+\alpha_{f}$, each with
a unique intercept $\alpha_{f}$ and corresponding gradient
$\alpha_{\nu}$. Assuming the emission from each GRB corresponds to a
combination of the high-latitude and afterglow models shown as the
solid and dashed lines (discussed above), we can parameterise each of
the GRBs with an intercept
$\alpha_{f}=(\alpha_{0}\alpha_{d}-3\beta/2)/(1-\beta/4)$ which is
given in Table~\ref{table4}. If objects have $\alpha_{f} \sim 0$ they
conform to the prediction of the afterglow model and if $\alpha_{f}
\sim 2$ they conform to the prediction of the high latitude
model. Thus $\alpha_{f}$ serves as a measure of the combination of
these 2 components.

We might expect those GRBs which have smaller $\alpha_f$ (more
afterglow dominated) to be more likely to have an early optical
detection. Using UVOT data\footnote{
{http:$//$swift.gsfc.nasa.gov$/$docs$/$swift$/$archive$/$grb\_table$/$}}
to quantify the early optical brightness, the data support such a
relationship. Of our sample, 33 GRBs have UVOT observations in the
first 10 minutes, of which 11 are detected in V. All of these 11 have
$\alpha_f < 0.85$. In contrast, of the 22 GRBs with upper limits
(typically V $> 19$), 19 have $\alpha_f > 0.85$. The likelihood of a
UVOT detection also correlates with $\alpha_0\alpha_d$, such that GRBs
with $\alpha_0\alpha_d < 2$ are four times more likely to have been
detected.

It is clear from Fig.~\ref{figure13} that the decay index,
$\alpha_0\alpha_d$, correlates with the strength of flares and humps,
i.e. the location of a burst in this figure depends on its location in
Fig.~\ref{figure10}. The bursts with the most significant humps do not
have large X-ray flares but they do have steep decays and straddle the
high latitude line in Fig.~\ref{figure13}. The bursts with no
significant flares or humps all lie below the high latitude line in
the bottom-left part of Fig.~\ref{figure13}.  In the context of the
ISM afterglow model, for these GRBs the implied electron energy index,
$p \sim 2 - 2.5$. For the bursts clustering around $\alpha_0\alpha_d
\sim 2.5$ and $\beta \sim 1.5$, which appear close to the afterglow
lines at the mid-right of Fig.~\ref{figure13}, the implied $p > 3$.

\begin{figure}
\figurenum{14}
\centerline{ 
\includegraphics[scale=0.7,angle=-90]{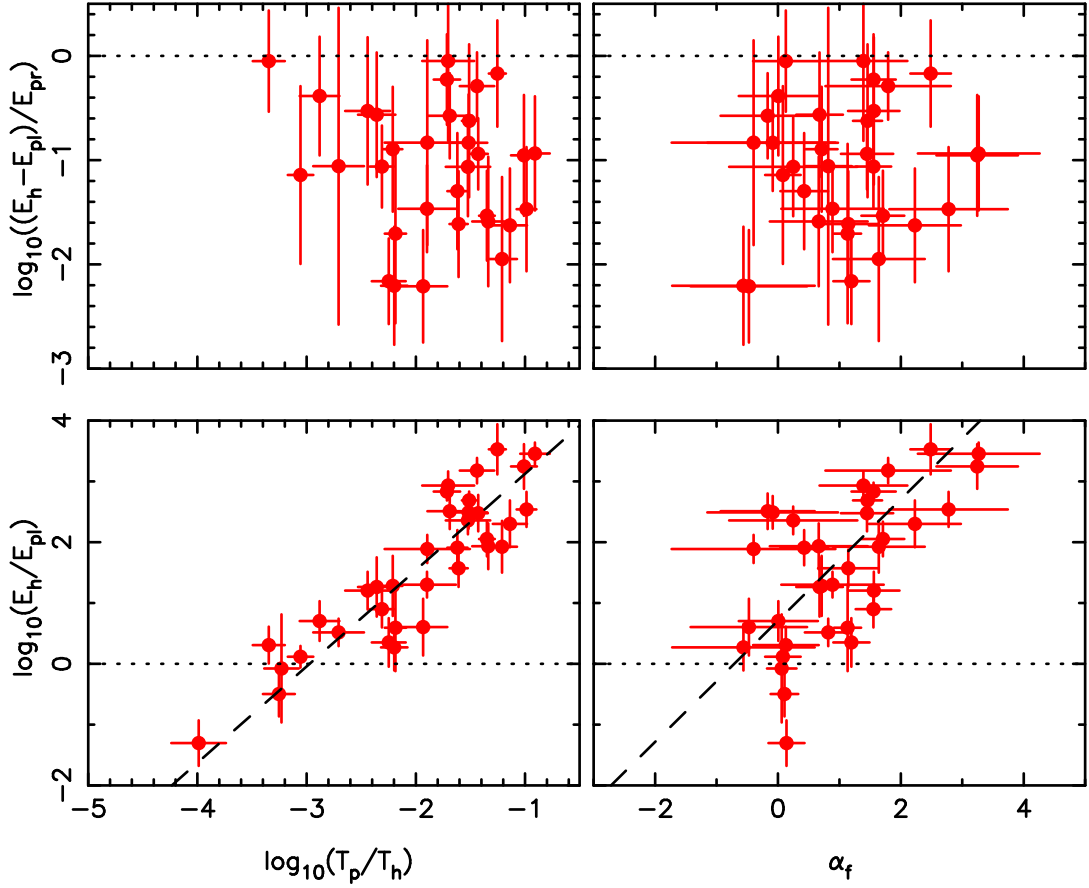}
}
\caption{
Bottom panels: The ratio of the size of the late hump emission to the
power law, $E_{h}/E_{pl}$, compared with T$_{p}$/T$_{h}$ and
$\alpha_{f}$.  The horizontal dotted line corresponds to equal fluence
in both.  Top panels: The ratio of the fluence in the late hump
emission to that of the prompt, $(E_{h}-E_{pl})/E_{pr}$, compared with
T$_{p}$/T$_{h}$ and $\alpha_{f}$.  The horizontal dotted line
corresponds to equal fluence in both.}
\label{figure14} \end{figure}

The late hump starts to appear at $\tau_{h}=3.5$, which corresponds to
a time T$_{h}=10^{(\tau_{h}+\tau_{d})/\alpha_{d}}$.  The ratio of the
prompt time (T$_{p}$) to this hump time is given by $\log_{10}({\rm
T}_{p}/{\rm T}_{h})=-1.8/\alpha_{d}$. We can integrate the light
curves to find the fluence for $t>$T$_{h}$, $E_{h}$, and compare this
with the fluence under the average power law for the same time
interval, $E_{pl}$. The ratio $E_{h}/E_{pl}$ is a measure of the size
of the hump with respect to the power law. The bottom left panel of
Fig.~\ref{figure14} shows the $\log_{10}$ of this ratio as a function
of $\log_{10}({\rm T}_{p}/{\rm T}_{h})$.  The horizontal dotted line
indicates the ratio value for which there is no excess late hump over
and above the power law decay. A simple explanation for the observed
correlation is that the weak hump component is always present at a
flux level a factor of $\sim10^{4}$ below the prompt emission and
lasts for $\sim1000\times {\rm T}_{p}$ seconds, but it is only
detected if the initial decay is fast with $\alpha_{0}\alpha_{d}>1.9$.
If the decay from the prompt phase is slow then, by the time the flux
has dropped to the appropriate level, the hump component has faded
away.

The bottom right-hand panel of Fig.~\ref{figure14} shows the
$\log_{10}$ of the fluence ratio $E_{h}/E_{pl}$ plotted
vs. $\alpha_{f}$. There is a significant correlation (coefficient
0.60, significant at $>99.9$\%) between the two indicating that the
late hump is more visible when the decay conforms to the high latitude
model ($\alpha_{f}\ge2$) but becomes obscured if an afterglow
component dominates.

We can also integrate the light curves for $t<$T$_{p}$ to find the
fluence of the prompt emission, $E_{pr}$. The ratio
$(E_{h}-E_{pl})/E_{pr}$ is then a measure of the fluence in the excess
late hump relative to the prompt fluence.  The upper panels of
Fig.~\ref{figure14} show $\log_{10}$ of this ratio with respect to
$\log_{10}({\rm T}_{p}/{\rm T}_{h})$ and $\alpha_{f}$.  There is no
correlation between the fluence in the excess late hump and temporal
decay index or the dominance of the high latitude emission over the
afterglow.  The GRBs with the largest $\alpha_{f}$ values (weak or no
afterglow component present) have the highest $\Delta_{H}$ and
$E_{h}/E_{pl}$ values but not the maximum $(E_{h}-E_{pl})/E_{pr}$. The
fluence in the late hump is not correlated with the presence of
afterglow nor is it correlated with the prompt fluence.

The horizontal dotted line in the top panels of Fig.~\ref{figure14}
indicates the level at which the late hump fluence is equal to the
prompt fluence. It is interesting that the maximum late hump fluence
is commensurate with the prompt fluence, suggestive of some kind of
equipartition in energy between these emission phases.  Previous
analyses of the hump have proposed that it is not consistent with an
afterglow component, but is consistent with refreshing of the forward
shock either by continued activity from the central engine or a range
in initial Lorentz factor of the ejecta (Nousek et al. 2006; Zhang et
al. 2006). At the end of energy injection (i.e. the end of the hump)
there will be a break in the decay index, the magnitude of which
depends on the injection mechanism (Nousek et al. 2006). After this
period the emission is presumably dominated by the afterglow
again. Another change in decay index will occur at the jet break. The
longest duration humps last almost a day, similar to the timescale on
which some jet breaks have been seen (Frail et al. 2001). This raises
the possibility that some light curve breaks described as a jet break
may be due to the end of energy injection.

The X-ray flares may also be due to the central engine, particularly
where the flare fluence is high (e.g. King et al. 2005).  In only two
cases --- GRB050502B and GRB050820A --- is there a clear indication
that the large, late flare appears to have caused an offset in the
later decay (i.e.the flux decays after the flare but does not rejoin
the previous decay). We note that Falcone et al. (2006) reach a
different conclusion for GRB050502B using XRT data alone. GRB050820A
is discussed in detail in Osborne et al. (2006).

\begin{figure}
\figurenum{15}
\centerline{ 
\includegraphics[scale=0.7,angle=-90]{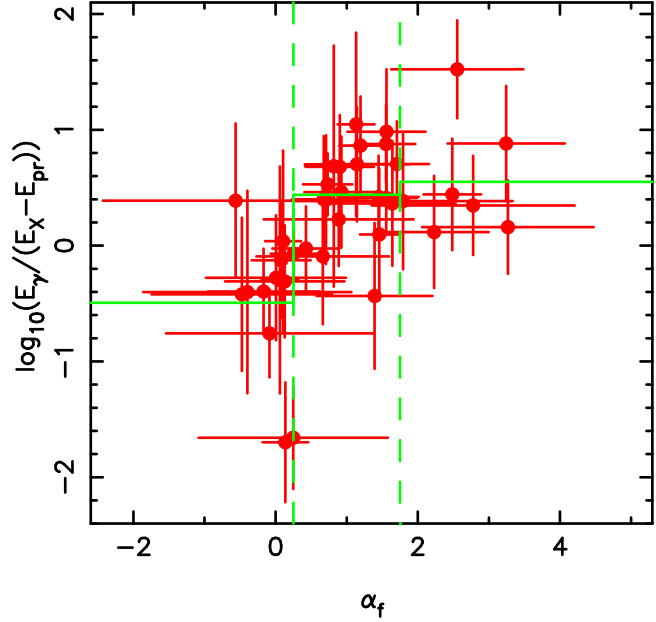}
}
\caption{
Comparison of the hard prompt fluence, E$_{\gamma}$
(15--150 keV) for $t<{\rm T}_{p}$, with the soft X-ray decay fluence,
E$_{X}-{\rm E}_{pr}$ (0.3--10 keV) for $t>{\rm T}_{p}$. The vertical
dashed lines correspond to $\alpha_f = 0.25$ and 1.75 respectively
(see text). }
\label{figure15} \end{figure}

The analysis described above has considered just the soft X-ray band,
0.3--10 keV (i.e. the XRT bandpass). We expect the prompt emission to
dominate in the hard (BAT) band, 15--150 keV, while the afterglow is
predominantly soft X-ray.  Fig.~\ref{figure15} shows the ratio of the
hard (15--150 keV) fluence for $t<$T$_{p}$, E$_{\gamma}$, with the
soft (0.3--10 keV) decay fluence for $t>{\rm T}_{p}$, $R_{E}={\rm
E}_{\gamma}/({\rm E}_{X}-{\rm E}_{pr})$, where E$_X$ is the total soft
fluence and E$_{pr}$ is the prompt fluence as before. For GRBs with
$\alpha_{f}>1.75$, $R_{E}=2.36\pm0.93$ while for $\alpha_{f}<0.25$
$R_{E}=0.22\pm0.13$. Thus, GRBs with smaller $\alpha_f$ have a greater
fraction of their energy emitted at $t>$T$_p$.

\begin{figure}
\figurenum{16}
\centerline{ 
\includegraphics[scale=0.7,angle=-90]{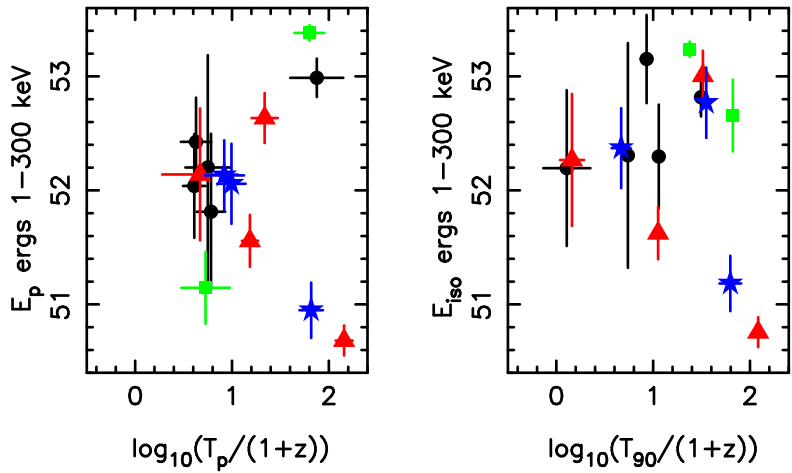}
}
\caption{
Left-hand panel: Isotropic equivalent energy in the hard
prompt emission, E$_{p}$ (1--300 keV) for $t<{\rm T}_{p}$,
calculated for those bursts for which we have a redshift. 
 Right-hand panel: Corresponding isotropic equivalent energy in the 
1--300 keV band derived over T$_{90}$.
Symbols as in Fig.~\ref{figure10}.}
\label{figure16} \end{figure}

For the GRBs with a known redshift we can use the burst duration,
T$_{p}$, to estimate the isotropic equivalent gamma-ray energy,
E$_{p}$ (1--300) keV, released in the rest-frame 1--300 keV band
during the prompt phase, $t<{\rm T}_{p}$. We have used a rest-frame
band similar to that of the BAT to avoid having to extrapolate over a
large energy range with a resultant more uncertain k-correction.  We
assume a cosmology with H$_0 = 71$ km s$^{-1}$ Mpc$^{-1}$, $\Lambda =
0.27$ and $\Omega = 0.73$.  These energy estimates are shown in the
left-hand panel of Fig.~\ref{figure16}, plotted against the log of the
rest-frame duration of the burst, $\log_{10}({\rm T}_{p}/(1+z))$. Most
bursts cluster in the center with similar durations and
luminosities. GRB050724 and GRB050803 lie towards the bottom
right. GRB050724 is a short burst with a long soft X-ray tail
(Barthelmy et al. 2005b) which causes a large T$_p$. It has a lower
luminosity relative to its duration than the long bursts plotted in
Fig.~\ref{figure16}. A low luminosity for GRB050724 is consistent with
other short bursts (Fox et al. 2005). GRB050803 has an uncertain
redshift as no clear optical transient was found. The redshift used
here ($z=0.422$) is that of a star-forming galaxy in the XRT error box
(Bloom et al. 2005). Either the redshift is under-estimated or this is
also an under-luminous burst.

The derived luminosity clearly depends on both the duration and the
bandpass over which the original fluence (to be converted into
luminosity) is obtained. To illustrate this, we also show in
Fig.~\ref{figure16} the isotropic equivalent gamma-ray energy,
E$_{iso}$ (1--300) keV, released in the rest-frame 1--300 keV band
during T$_{90}$.

\section{Conclusions} 

We have analysed data for the 40 GRBs observed by \swift\ prior to
2005 October 1 for which XRT observations began within 10 minutes of
the BAT trigger. We have combined data from the BAT and the XRT to
investigate the form of the X-ray emission (0.3--10 keV) during the
first few hours following the burst.

The initial XRT spectral index is slightly steeper than that seen in
the BAT, showing that spectral evolution occurs as the GRB
ages. Combining the BAT and XRT data, the raw light curves show that
the initial X-ray emission seen in the XRT is consistent with being a
continuation of the emission seen by the BAT.  Some two-thirds of the
GRBs display a light curve which shows a steeply declining component
that breaks to a shallower decay rate -- the late emission hump --
usually within an hour of the trigger. The remaining bursts decline
fairly continuously. At least half of the GRBs in the sample display
late X-ray flares, probably due to continued central engine activity,
but in only a few GRBs does the fluence in the flares rival that of
the burst.

To investigate the early X-ray emission, we used an automatic fitting
procedure to align the light curves. Allowing for flares, this
procedure works well for the entire sample. The resultant light curves
display a prompt phase, mostly observed by the BAT, followed by a
decline. The light curve can be described by an exponential which
relaxes into a power law whose decay rate varies considerably from
burst to burst. The transition time between the exponential and power
law provides a well determined measure of the burst duration.

Comparing the temporal and spectral indices of the power law decline,
the distribution is consistent with a simple model in which the early
emission is a combination of emission from the central engine
(parameterised by high latitude emission; Kumar \& Panaitescu (2000))
and afterglow. Those GRBs in which the afterglow is weak early on
decay fast during the power law phase and their X-ray light curves are
consistent with the high latitude model. Some are dominated by
afterglow, while the majority require a significant contribution from
both components. The likelihood of an early optical detection strongly
correlates with the strength of the X-ray afterglow component.

The late emission hump component may be present in all objects but can
be masked by a strong afterglow component. The late hump can last for
many tens of thousands of seconds and may also be due to continued
central engine activity (Nousek et al. 2006; Zhang et
al. 2006). Interestingly, the strongest humps seen have a total
fluence which matches that of the prompt phase.  There appears to be a
correlation such that bursts with the most visible humps do not have
strong, late X-ray flares.

\paragraph*{Acknowledgments}

MRG, JPO, KP, APB, SV, ER and OG gratefully acknowledge funding
through the PPARC.  This work is also supported at Pennsylvania State
University (PSU) by NASA contract NAS5-00136 and NASA grant 
NNG05GF43G, and at the Osservatorio
Astronomico di Brera (OAB) by funding from ASI on grant number
I/R/039/04. We gratefully acknowledge the contributions of numerous
colleagues at Leicester University, PSU, OAB, Goddard Space Flight
Center and our sub-contractors, who helped make the BAT and XRT
possible.

\clearpage

\begin{deluxetable}{lrrrclc}
\tablecaption{GRB sample and BAT data}
\tablecolumns{7}
\tablehead{
\colhead{GRB} & 
\colhead{T$_{90}$} &
\colhead{T$_{50}$} & 
\colhead{$\beta_b $} &
\colhead{BAT mean flux} &
\colhead{Redshift} &
\colhead{Ref.\tablenotemark{\dagger}}
\\
\colhead{} &
\colhead{} &
\colhead{} &
\colhead{} &
\colhead{(15--150 keV)} &
\colhead{} &
\colhead{}
\\
\colhead{} & 
\colhead{(s)} & 
\colhead{(s)} &
\colhead{} &
\colhead{($ 10^{-8}$ erg cm$^{-2}$ s$^{-1}$)} &
\colhead{} &
\colhead{}
}    
\startdata
050126 & 25.7 $\pm$ 0.3 & 13.7 $\pm$ 0.3 & 0.41 $\pm$ 0.15 & 3.15 $\pm$ 0.28 & 1.29  & 1 \\
050128 & 28.0 $\pm$ 0.2 & 8.0 $\pm$ 0.2 & 0.41 $\pm$ 0.08 & 17.1 $\pm$ 0.71 &  & \\
050219A & 23.5 $\pm$ 0.3 & 9.9 $\pm$ 0.2 & 0.35 $\pm$ 0.05 & 16.7 $\pm$ 0.51 &  & \\ 
050315 & 96.0 $\pm$ 0.5 & 24.7 $\pm$ 0.4 & 1.15 $\pm$ 0.09 & 3.23 $\pm$ 0.14 & 1.949  & 2 \\
050319 & 149.6 $\pm$ 0.7 & 58 $\pm$ 0.5 & 1.10 $\pm$ 0.20 & 0.54 $\pm$ 0.05 & 3.24  & 3 \\ 
050401 & 33.3 $\pm$ 2.0 & 25.8 $\pm$ 1.0 & 0.52 $\pm$ 0.07 & 22.3 $\pm$ 0.88 & 2.90  & 4 \\ 
050406 & 5.7 $\pm$ 0.2 & 2.4 $\pm$ 0.3 & 1.64 $\pm$ 0.47 & 1.35 $\pm$ 0.31 &  & \\
050412 & 24.1 $\pm$ 2.0 & 8.4 $\pm$ 1.0 & $-0.26$ $\pm$ 0.18 & 2.22 $\pm$ 0.22 &  & \\ 
050416A & 2.4 $\pm$ 0.3 & 0.7 $\pm$ 0.2 & 2.20 $\pm$ 0.25 & 13.76 $\pm$ 1.31 & 0.6535  & 5 \\ 
050421 & 10.3 $\pm$ 0.2 & 5.1 $\pm$ 0.2 & 0.64 $\pm$ 0.46 & 1.04 $\pm$ 0.29 &  & \\ 
050422 & 59.2 $\pm$ 0.3 & 42.8 $\pm$ 0.3 & 0.54 $\pm$ 0.21 & 0.97 $\pm$ 0.12 &  & \\ 
050502B & 17.5 $\pm$ 0.2 & 9.8 $\pm$ 0.2 & 0.64 $\pm$ 0.15 & 2.57 $\pm$ 0.23 &  & \\ 
050525A & 8.8 $\pm$ 0.5 & 5.2 $\pm$ 0.4 & 0.83 $\pm$ 0.02 & 175.9 $\pm$ 2.32 & 0.606  & 6 \\ 
050607 & 26.5 $\pm$ 0.2 & 13.7 $\pm$ 0.2 & 0.97 $\pm$ 0.17 & 2.15 $\pm$ 0.20 &  & \\ 
050712 & 48.4 $\pm$ 2.0 & 25.6 $\pm$ 1.0 & 0.50 $\pm$ 0.19 & 1.96 $\pm$ 0.22 &  & \\ 
050713A & 128.8 $\pm$ 10 & 11.4 $\pm$ 10 & 0.55 $\pm$ 0.07 & 3.81 $\pm$ 0.14 &  & \\ 
050713B & 131.0 $\pm$ 3.0 & 44.0 $\pm$ 0.7 & 0.53 $\pm$ 0.15 & 3.21 $\pm$ 0.27 &  & \\ 
050714B & 46.4 $\pm$ 0.4 & 21.4 $\pm$ 0.3 & 1.70 $\pm$ 0.41 & 1.13 $\pm$ 0.21 &  & \\ 
050716 & 69.4 $\pm$ 1.0 & 36.0 $\pm$ 0.5 & 0.47 $\pm$ 0.06 & 8.76 $\pm$ 0.30 &  & \\ 
050717 & 67.2 $\pm$ 2.0 & 24.8 $\pm$ 1.0 & 0.36 $\pm$ 0.05 & 8.63 $\pm$ 0.20 &  & \\
050721 & 39.2 $\pm$ 0.5 & 15.0 $\pm$ 0.5 & 0.78 $\pm$ 0.12 & 7.11 $\pm$ 0.48 &  & \\
050724\tablenotemark{*} & 152.5 $\pm$ 9.0 & 84.6 $\pm$ 3.0 & 1.17 $\pm$ 0.26 & 0.74 $\pm$ 0.09 & 0.257  & 7 \\
050726 & 33.9 $\pm$ 0.2 & 15.0 $\pm$ 0.5 & 0.01 $\pm$ 0.17 & 4.84 $\pm$ 0.43 &  & \\
050730 & 155.0 $\pm$ 2.1 & 63.0 $\pm$ 1.6 & 0.52 $\pm$ 0.11 & 1.48 $\pm$ 0.09 & 3.97  & 8 \\
050801 & 20.0 $\pm$ 3.0 & 6.0 $\pm$ 1.0 & 1.03 $\pm$ 0.24 & 1.52 $\pm$ 0.22 &  & \\
050802 & 30.9 $\pm$ 1.0 & 9.3 $\pm$ 0.5 & 0.66 $\pm$ 0.15 & 6.08 $\pm$ 0.52 & 1.71 & 9 \\
050803 & 89.0 $\pm$ 10 & 40.0 $\pm$ 5.0 & 0.47 $\pm$ 0.11 & 2.38 
$\pm$ 0.15 & (0.422)\tablenotemark{\ddagger} & 10 \\
050813\tablenotemark{*} & 0.58 $\pm$ 0.1 & 0.26 $\pm$ 0.1 & 0.37 $\pm$ 0.37 & 8.52 $\pm$ 0.20&  & \\
050814 & 144.0 $\pm$ 3.0 & 56.0 $\pm$ 1.2 & 0.98 $\pm$ 0.19 & 1.22 $\pm$ 0.13 &  & \\
050819 & 35.8 $\pm$ 4.0 & 16.4 $\pm$ 1.0 & 1.56 $\pm$ 0.21 & 0.91 $\pm$ 0.14 &  & \\
050820A & 240 $\pm$ 5.0 & 221 $\pm$ 5.0 & 0.24 $\pm$ 0.07 & 1.57 $\pm$ 0.11 & 2.612 & 11 \\
050822 & 105 $\pm$ 2.0 & 43.8 $\pm$ 1.0 & 1.53 $\pm$ 0.09 & 2.54 $\pm$ 0.16 &  & \\
050826 & 35.3 $\pm$ 8.0 & 14.8 $\pm$ 2.0 & 0.10 $\pm$ 0.28 & 1.18 $\pm$ 0.19 &  & \\
050904 & 173.2 $\pm$ 10 & 53.3 $\pm$ 5.0 & 0.38 $\pm$ 0.04 & 2.74 $\pm$ 0.10 & 6.29 & 12 \\
050908 & 20.3 $\pm$ 2.0 & 6.7 $\pm$ 1.0 & 0.91 $\pm$ 0.11 & 2.41 $\pm$ 0.26 & 3.35  & 13 \\
050915A & 41.8 $\pm$ 3.0 & 14.2 $\pm$ 1.0 & 0.37 $\pm$ 0.11 & 1.79 $\pm$ 0.18 &  & \\
050915B & 39.5 $\pm$ 1.0 & 21.3 $\pm$ 0.6 & 0.89 $\pm$ 0.06 & 8.61 $\pm$ 0.34 &  & \\
050916 & 92.1 $\pm$ 10 & 32.0 $\pm$ 4.0 & 0.83 $\pm$ 0.32 & 1.19 $\pm$ 0.16 &  & \\
050922B & 150 $\pm$ 15 & 25.0 $\pm$ 5.0 & 1.11 $\pm$ 0.16 & 1.25 $\pm$ 0.18 &  & \\
050922C & 4.1 $\pm$ 1.0 & 1.4 $\pm$ 0.5 & 0.34 $\pm$ 0.03 & 35.7 $\pm$ 1.05 & 2.198  & 14 \\
\enddata
\tablenotetext{*}{Would have appeared as a short burst to BATSE.}
\tablenotetext{\ddagger}{Uncertain identification. Redshift given is that 
of a star-forming galaxy in XRT error box.}
\tablenotetext{\dagger}{Redshift references. 
1: Berger, Cenko, \& Kulkarni (2005). 
2: Kelson \& Berger (2005).
3: Fynbo et al. (2005a).
4: Fynbo et al. (2005b).
5: Cenko et al. (2005).
6: Foley, Chen \& Bloom (2005).
7: Prochaska et al. (2005a).
8: Chen et al. (2005).
9. Fynbo et al. (2005c).
10. Bloom et al. (2005).
11: Prochaska et al. (2005b).
12: Kawai et al. (2005).
13: Fugazza et al. (2005).
14: Jakobsson et al. (2005).
} 
\label{table1}
\end{deluxetable}

\clearpage

\begin{deluxetable}{lccl}
\tablecaption{BAT fits for those bursts better-fitted with a cutoff power law}
\tablecolumns{4}
\tablehead{
\colhead{GRB} &  
\colhead{BAT $\beta_{bc} $} &
\colhead{E$_{\rm cut}$} &
\colhead{$\Delta \chi^{2}$\tablenotemark{\dagger}}
\\
\colhead{} &
\colhead{} &
\colhead{keV} &
\colhead{}
}    
\startdata
050128  & $-0.53^{+0.35}_{-0.37}$ & $65.4^{+39.6}_{-18.6}$ & 23.8 \\
050219A & $-1.03^{+0.28}_{-0.30}$ & $43.2^{+10.8}_{-7.60}$ & 88.3 \\
050525A & $-0.17^{+0.12}_{-0.12}$ & $63.8^{+8.30}_{-6.70}$ & 246.6 \\
050716  & $-0.17^{+0.27}_{-0.29}$ & $89.1^{+61.6}_{-27.4}$ & 18.0 \\
\enddata
\tablenotetext{\dagger}{Improvement in $\chi^{2}$ for one 
degree of freedom.}
\label{table2}
\end{deluxetable}

\clearpage

\begin{deluxetable}{lccclccc}
\tablecaption{XRT spectral and temporal fits}
\tablecolumns{8}
\tablehead{
\colhead{GRB} & 
\colhead{XRT start} &
\colhead{XRT $\beta_x$} &
\colhead{Galactic} &
\colhead{Intrinsic} &
\colhead{$\alpha_1$} &
\colhead{$t_{\rm break}$} &
\colhead{$\alpha_2$}  
\\
\colhead{} & 
\colhead{(s)} &
\colhead{} &
\multicolumn{2}{c}{N$_{\rm H}$ ($10^{20}$ cm$^{-2}$)} &
\colhead{} &
\colhead{(s)} &
\colhead{}  

}    
\startdata
050126& $127$ & $1.59 \pm 0.38$ & $5.30$ & $$ & $2.52^{+0.50}_{-0.22}$ & $424^{+561}_{-120}$ & $1.00^{+0.17}_{-0.26}$ \\
050128& $227$ & $0.85 \pm 0.12$ & $4.80$ & $7.67^{+2.06}_{-1.85}$ & $0.66^{+0.10}_{-0.11}$ & $1724^{+937}_{-565}$ & $1.16^{+0.09}_{-0.08}$ \\
050219A& $92$ & $1.02 \pm 0.20$ & $8.50$ & $17.0^{+7.00}_{-6.40}$ & $3.17^{+0.24}_{-0.16}$ & $332^{+26}_{-22}$ & $0.75^{+0.09}_{-0.07}$ \\
050315& $83$  & $1.50 \pm 0.40$  & $4.30$ & $122^{+46.0}_{-38.0}\tablenotemark{*}$ & $5.30^{+0.50}_{-0.40}$ & $400^{+20}_{-20}$ & $0.71^{+0.04}_{-0.04}$ \\
050319& $211$ & $2.02 \pm 0.47$ & $1.10$ & $$ & $3.80^{+0.56}_{-0.56}$ & $424^{+38}_{-35}$ & $0.47^{+0.10}_{-0.10}$ \\
050401& $128$ & $0.98 \pm 0.05$ & $4.80$ & $170^{+37.0}_{-34.0}\tablenotemark{*}$ & $0.76^{+0.02}_{-0.02}$ & $5518^{+1149}_{-1043}$ & $1.31^{+0.05}_{-0.05}$ \\
050406& $84$  & $1.37 \pm 0.25$ & $3.00$ & $$ & $1.05^{+0.57}_{-0.51}$ & $$ & $$ \\
050412& $99$  & $0.26 \pm 0.32$ & $2.20$ & $$ & $1.81^{+0.57}_{-0.47}$ & $$ & $$ \\
050416A& $87$ & $0.80 \pm 0.29$ & $2.10$ & $46.8^{+31.7}_{-25.2}\tablenotemark{*}$ & $0.87^{+0.45}_{-0.45}$ & $400^{+221}_{-264}$ & $0.24^{+0.44}_{-0.24}$ \\
050421& $110$ & $0.27 \pm 0.37$ & $14.4$ & $61.0^{+42.0}_{-35.0}$ & $3.05^{+0.17}_{-0.15}$ & $$ & $$ \\
050422& $109$ & $2.33 \pm 0.60$ & $100$ & $$ & $5.31^{+0.66}_{-0.60}$ & $341^{+154}_{-72}$ & $0.59^{+0.20}_{-0.27}$ \\
050502B& $63$ & $0.81 \pm 0.28$ & $3.70$ & $$ & $1.35^{+0.31}_{-0.27}$ & $$ & $$ \\
050525A& $125$& $1.07 \pm 0.02$ & $9.00$ & $38.0^{+3.40}_{-3.00}\tablenotemark{*}$ & $0.98^{+0.05}_{-0.05}$ & $641^{+690}_{-123}$ & $1.39^{+0.09}_{-0.04}$ \\
050607& $84$  & $0.77 \pm 0.48$ & $14.0$ & $$ & $1.94^{+0.17}_{-0.17}$ & $1217^{+372}_{-276}$ & $0.54^{+0.05}_{-0.05}$ \\
050712& $166$ & $0.90 \pm 0.06$ & $13.0$ & $$ & $1.34^{+0.10}_{-0.10}$ & $3987^{+2576}_{-2064}$ & $0.76^{+0.10}_{-0.10}$ \\
050713A& $73$ & $1.30 \pm 0.07$ & $11.0$ & $42.0^{+3.20}_{-2.90}$ & $2.29^{+0.13}_{-0.14}$ & $321^{+107}_{-38}$ & $0.71^{+0.08}_{-0.17}$ \\
050713B& $136$& $0.70 \pm 0.11$ & $18.0$ & $20.0^{+4.90}_{-4.50}$ & $2.88^{+0.14}_{-0.13}$ & $540^{+47}_{-44}$ & $0.43^{+0.06}_{-0.06}$ \\
050714B& $151$& $4.50 \pm 0.70$ & $5.30$ & $63.0^{+13.4}_{-11.7}$ & $6.96^{+0.60}_{-0.60}$ & $366^{+38}_{-38}$ & $0.52^{+0.11}_{-0.11}$ \\
050716& $96$  & $0.33 \pm 0.03$ & $11.0$ & $$ & $2.09^{+0.04}_{-0.03}$ & $1700^{+434}_{-329}$ & $1.02^{+0.07}_{-0.08}$ \\
050717& $79$  & $0.63 \pm 0.11$ & $23.0$ & $31.0^{+6.50}_{-6.10}$ & $1.95^{+0.18}_{-0.11}$ & $318^{+115}_{-95}$ & $1.29^{+0.08}_{-0.08}$ \\
050721& $186$ & $0.74 \pm 0.15$ & $16.0$ & $19.0^{+6.40}_{-5.70}$ & $2.35^{+0.16}_{-0.16}$ & $380^{+30}_{-30}$ & $1.22^{+0.03}_{-0.03}$ \\
050724& $74$  & $0.95 \pm 0.07$ & $59\tablenotemark{\dagger}$ & $$ & $4.03^{+0.18}_{-0.15}$ & $508^{+120}_{-31}$ & $1.82^{+0.24}_{-0.28}$ \\
050726& $110$ & $0.94 \pm 0.07$ & $4.70$ & $$ & $0.97^{+0.30}_{-0.30}$ & $7754^{+1646}_{-2162}$ & $1.78^{+0.30}_{-0.25}$ \\
050730& $130$ & $0.33 \pm 0.08$ & $3.10$ & $140^{+37.0}_{-35.0}\tablenotemark{*}$ & $2.21^{+0.44}_{-0.44}$ & $222^{+25}_{-21}$ & $0.35^{+0.30}_{-0.20}$ \\
050801& $61$  & $0.72 \pm 0.54$ & $7.00$ & $$ & $0.96^{+0.04}_{-0.04}$ & $$ & $$ \\
050802& $289$ & $0.91 \pm 0.19$ & $1.80$ & $$ & $0.64^{+0.10}_{-0.09}$ & $6036^{+1667}_{-850}$ & $1.66^{+0.06}_{-0.06}$ \\
050803& $152$ & $0.71 \pm 0.16$ & $5.60$ & $20.0^{+8.60}_{-7.70}\tablenotemark{*}$ & $5.15^{+0.26}_{-0.26}$ & $272^{+12}_{-8}$ & $0.59^{+0.03}_{-0.04}$ \\
050813& $73$  & $2.42 \pm 0.89$ & $4.00$ & $$ & $2.00^{+0.61}_{-0.50}$ & $$ & $$  \\
050814& $138$ & $1.08 \pm 0.08$ & $2.60$ & $2.20^{+1.35}_{-1.27}$ & $3.03^{+0.09}_{-0.08}$ & $1039^{+130}_{-109}$ & $0.66^{+0.08}_{-0.08}$ \\
050819& $141$ & $1.18 \pm 0.23$ & $4.70$ & $$ & $4.39^{+0.50}_{-0.50}$ & $269^{+40}_{-21}$ & $2.09^{+0.53}_{-0.57}$ \\
050820A&$80$  & $0.87 \pm 0.09$ & $4.60$ & $$ & $2.22^{+0.23}_{-0.22}$ & $2060^{+1850}_{-1850}$ & $1.13^{+0.02}_{-0.02}$ \\
050822& $96$  & $1.60 \pm 0.06$ & $2.30$ & $13.0^{+1.00}_{-0.90}$ & $2.81^{+0.25}_{-0.25}$ & $279^{+117}_{-117}$ & $0.39^{+0.12}_{-0.14}$ \\
050826& $109$ & $1.27 \pm 0.47$ & $22.0$ & $65.0^{+51.8}_{-44.2}$ & $1.19^{+0.07}_{-0.05}$ & & \\
050904& $161$ & $0.44 \pm 0.04$ & $4.90$ & $355^{+10.0}_{-97.0}\tablenotemark{*}$ & $1.81^{+0.06}_{-0.06}$ & & \\
050908& $106$ & $2.35 \pm 0.27$ & $2.10$ & $$ & $1.04^{+0.07}_{-0.13}$ & & \\
050915A& $87$ & $1.12 \pm 0.34$ & $1.90$ & $15.0^{+9.00}_{-9.00}$ & $2.87^{+1.40}_{-1.40}$ & $144^{+21}_{-30}$ & $0.89^{+0.03}_{-0.03}$ \\
050915B& $136$ & $1.45 \pm 0.10$ & $30.0$ & $$ & $5.21^{+0.16}_{-0.16}$ & $437^{+15}_{-15}$ & $0.66^{+0.08}_{-0.08}$ \\
050916& $210$ & $0.77 \pm 0.84$ & $122$ & $$ & $0.79^{+0.08}_{-0.08}$ \\
050922B& $342$ & $1.64 \pm 0.08$ & $3.40$ & $12.0^{+1.70}_{-1.50}$ & $3.04^{+0.18}_{-0.18}$ & $2971^{+388}_{-388}$ & $0.30^{+0.08}_{-0.08}$ \\
050922C& $108$ & $1.10 \pm 0.09$ & $5.80$ & $$ & $1.19^{+0.02}_{-0.02}$ \\
\enddata
\tablenotetext{*}{In rest-frame.}
\tablenotetext{\dagger }{Vaughan et al. (2006b)}
\label{table3}
\end{deluxetable}

\clearpage

\begin{deluxetable}{lrrrrr}
\tablecaption{
Best fit parameters derived from the average X-ray
decay curve.} 
\tablecolumns{6}
\tablehead{
\colhead{GRB} &  
\colhead{$\alpha_0\alpha_d$} &
\colhead{$\Delta_F$} &
\colhead{$\Delta_H$} &
\colhead{T$_p$} &
\colhead{$\alpha_f$}
}    
\startdata
 050126&      2.36 $\pm$   0.20 &  $-$0.03 $\pm$   0.31&   2.15 $\pm$   0.31 &   35.1 $\pm$    2.9&   1.14 $\pm$   0.49\\
 050128&      1.13 $\pm$   0.16 &   0.01 $\pm$   0.12& &   38.7 $\pm$    5.3&   0.93 $\pm$   0.34\\
 050219A&     1.55 $\pm$   0.32 &   0.03 $\pm$   0.22&   1.06 $\pm$   0.31&   39.8 $\pm$    8.2&   1.56 $\pm$   0.42\\
 050315&      4.17 $\pm$   0.56 &   0.08 $\pm$   0.28&   9.72 $\pm$   0.18 &   64.2 $\pm$    8.7&   3.27 $\pm$   0.99\\
 050319&      2.49 $\pm$   0.51 &   0.40 $\pm$   0.22&   3.00 $\pm$   0.23 &   35.3 $\pm$    7.2&   0.25 $\pm$   1.04\\
 050401&      1.17 $\pm$   0.19 &   0.08 $\pm$   0.15&   0.01 $\pm$   0.89 &   16.6 $\pm$    2.6&   0.06 $\pm$   0.24\\
 050406&      1.96 $\pm$   0.44 &   0.48 $\pm$   0.32&   0.73 $\pm$   0.47 &    7.3 $\pm$    1.6&  $-$0.47 $\pm$   0.95\\
 050412&      0.90 $\pm$   0.14 &   0.10 $\pm$   0.28& &   12.7 $\pm$    1.9&   0.90 $\pm$   0.31\\
 050416A&     2.00 $\pm$   0.78 &  $-$0.32 $\pm$   0.61&   2.29 $\pm$   0.23 &    7.7 $\pm$    3.0&  $-$0.40 $\pm$   1.33\\
 050421&      2.95 $\pm$   0.22 &   0.03 $\pm$   0.38& &  172 $\pm$   13&   2.56 $\pm$   0.60\\
 050422&      3.84 $\pm$   0.35 &   0.00 $\pm$   0.23&   7.22 $\pm$   0.29 &   46.5 $\pm$    4.2&   2.70 $\pm$   1.49\\
 050502B&     2.23 $\pm$   0.53 &   2.08 $\pm$   0.40&   3.92 $\pm$   0.23 &   32.5 $\pm$    7.8&   1.39 $\pm$   0.71\\
 050525A&     1.40 $\pm$   0.33 &   0.29 $\pm$   0.24&   0.51 $\pm$   0.23 &    9.0 $\pm$    2.1&   0.82 $\pm$   0.38\\
 050607&      2.00 $\pm$   0.53 &   0.49 $\pm$   0.23&   1.77 $\pm$   0.21 &   26.6 $\pm$    7.0&   0.89 $\pm$   0.83\\
 050712&      1.61 $\pm$   0.28 &   0.04 $\pm$   0.12&   1.22 $\pm$   0.49 &  131 $\pm$   22&   0.68 $\pm$   0.38\\
 050713A&     2.51 $\pm$   0.17 &   1.15 $\pm$   0.12&   3.98 $\pm$   0.14 &   13.9 $\pm$    0.9&   1.46 $\pm$   0.24\\
 050713B&     3.03 $\pm$   0.24 &   0.03 $\pm$   0.18&   5.54 $\pm$   0.41 &  143 $\pm$   11&   2.49 $\pm$   0.33\\
 050714B&     2.50 $\pm$   0.43 &   0.20 $\pm$   0.19&   3.09 $\pm$   0.27 &   96 $\pm$   17&  $-$0.09 $\pm$   1.06\\
 050716&      1.64 $\pm$   0.18 &   0.09 $\pm$   0.08&   0.69 $\pm$   0.30 &  124. $\pm$   14&   1.55 $\pm$   0.29\\
 050717&      1.73 $\pm$   0.17 &   0.02 $\pm$   0.11&   0.55 $\pm$   0.71 &   58.4 $\pm$    5.7&   1.13 $\pm$   0.22\\
 050721&      1.72 $\pm$   0.15 &  $-$0.04 $\pm$   0.12&   1.02 $\pm$   0.49 &   56.0 $\pm$    5.0&   0.71 $\pm$   0.26\\
 050724&      2.66 $\pm$   0.23 &   0.04 $\pm$   0.14&   3.09 $\pm$   0.30 &  181 $\pm$   16&   1.45 $\pm$   0.43\\
 050726&      1.36 $\pm$   0.20 &   0.06 $\pm$   0.10& &   46.5 $\pm$    7.0&   0.73 $\pm$   0.28\\
 050730&      0.87 $\pm$   0.24 &   0.16 $\pm$   0.12& &  373 $\pm$  102&   0.25 $\pm$   0.29\\
 050801&      1.32 $\pm$   0.24 &  $-$0.02 $\pm$   0.23&   0.44 $\pm$   0.33 &   10.3 $\pm$    1.9&   0.01 $\pm$   0.64\\
 050802&      1.24 $\pm$   0.15 &   0.12 $\pm$   0.10&   0.29 $\pm$   0.18 &   11.1 $\pm$    1.3&   0.08 $\pm$   0.29\\
 050803&      2.21 $\pm$   0.27 &   0.43 $\pm$   0.15&   2.56 $\pm$   0.15 &   94 $\pm$   11&   1.56 $\pm$   0.36\\
 050813&      1.73 $\pm$   0.21 &   0.21 $\pm$   0.42&   0.30 $\pm$   0.38 &    0.7 $\pm$    0.1&  $-$0.56 $\pm$   1.16\\
 050814&      2.81 $\pm$   0.20 &   0.03 $\pm$   0.10&   2.55 $\pm$   0.29 &  131 $\pm$    9.5&   1.71 $\pm$   0.35\\
 050819&      3.13 $\pm$   0.43 &  $-$0.01 $\pm$   0.27&   3.69 $\pm$   0.43 &   55.6 $\pm$    7.6&   1.64 $\pm$   0.75\\
 050820A&     0.95 $\pm$   0.24 &   0.66 $\pm$   0.17&   0.17 $\pm$   0.38 &   19.3 $\pm$    4.8&   0.14 $\pm$   0.29\\
 050822&      2.24 $\pm$   0.46 &   0.45 $\pm$   0.08&   2.98 $\pm$   0.29 &  140 $\pm$   29&  $-$0.17 $\pm$   0.77\\
 050826&      1.13 $\pm$   0.17 &   0.17 $\pm$   0.26&   0.29 $\pm$   0.30 &   15.5 $\pm$    2.3&   0.13 $\pm$   0.53\\
 050904&      1.69 $\pm$   0.26 &   0.46 $\pm$   0.13&   0.59 $\pm$   0.40 &  460 $\pm$   71&   1.20 $\pm$   0.30\\
 050908&      2.84 $\pm$   0.42  &   0.75 $\pm$   0.28&   3.29 $\pm$   0.38 &   42.9 $\pm$    6.3&   0.66 $\pm$   0.80\\
 050915A&     3.76 $\pm$   0.45 &   0.31 $\pm$   0.21&   6.40 $\pm$   0.37 &   79.9 $\pm$    9.5&   3.24 $\pm$   0.66\\
 050915B&     3.33 $\pm$   0.52 &  $-$0.01 $\pm$   0.19&   3.93 $\pm$   0.39 &   82 $\pm$   13&   2.23 $\pm$   0.75\\
 050916&      2.63 $\pm$   0.41 &  $-$0.11 $\pm$   0.54&   4.06 $\pm$   0.21 &   84 $\pm$   13&   1.79 $\pm$   1.02\\
 050922B&     2.34 $\pm$   0.22 &   0.16 $\pm$   0.08&   2.53 $\pm$   0.29 &  143 $\pm$   13&   0.43 $\pm$   0.39\\
 050922C&     1.17 $\pm$   0.17 &   0.03 $\pm$   0.09&  $-$0.18 $\pm$   0.37 &   19.4 $\pm$    2.8&   0.10 $\pm$   0.23\\
\enddata
\label{table4}
\end{deluxetable}

\clearpage

\begin{deluxetable}{lcr}
\tablecaption{Summary of significant correlations}
\tablecolumns{3}
\tablehead{
\colhead{Variables\tablenotemark{*}} &  
\colhead{Spearman Corr. Coeff.} &
\colhead{Significance (\%)}
}    
\startdata
$\alpha_1$, $\beta_x$ & 0.30 & 95 \\
($\alpha_1 - \beta_x$), $\alpha_1$ & 0.89 & $\gg 99.9$ \\
$\alpha_0\alpha_d$, $\beta$ (entire sample) & 0.53 & $99$ \\
$\alpha_0\alpha_d$, $\beta$ (GRBs below high lat. line in Fig.~\ref{figure13}) & 0.66 & $\gg 99.9$ \\
$E_h/E_{pl}$, $\alpha_f$ & 0.60 & $>99.9$ \\
\enddata
\tablenotetext{*}{Defined in text.}
\label{table5}
\end{deluxetable}
\clearpage

\clearpage

\clearpage

\centerline{
\includegraphics[]{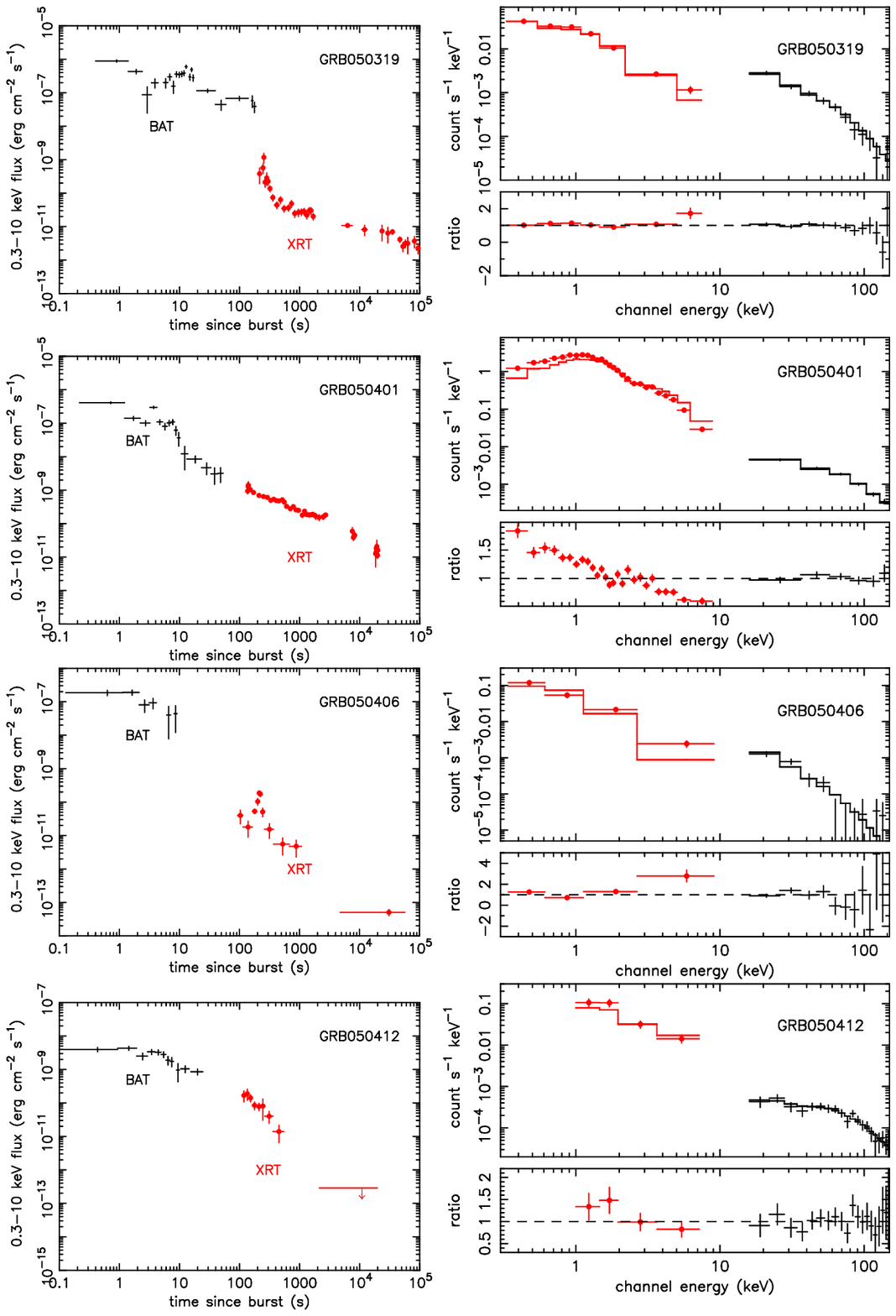}
}
\centerline{Fig. 2. --- continued.}

\clearpage

\centerline{
\includegraphics[]{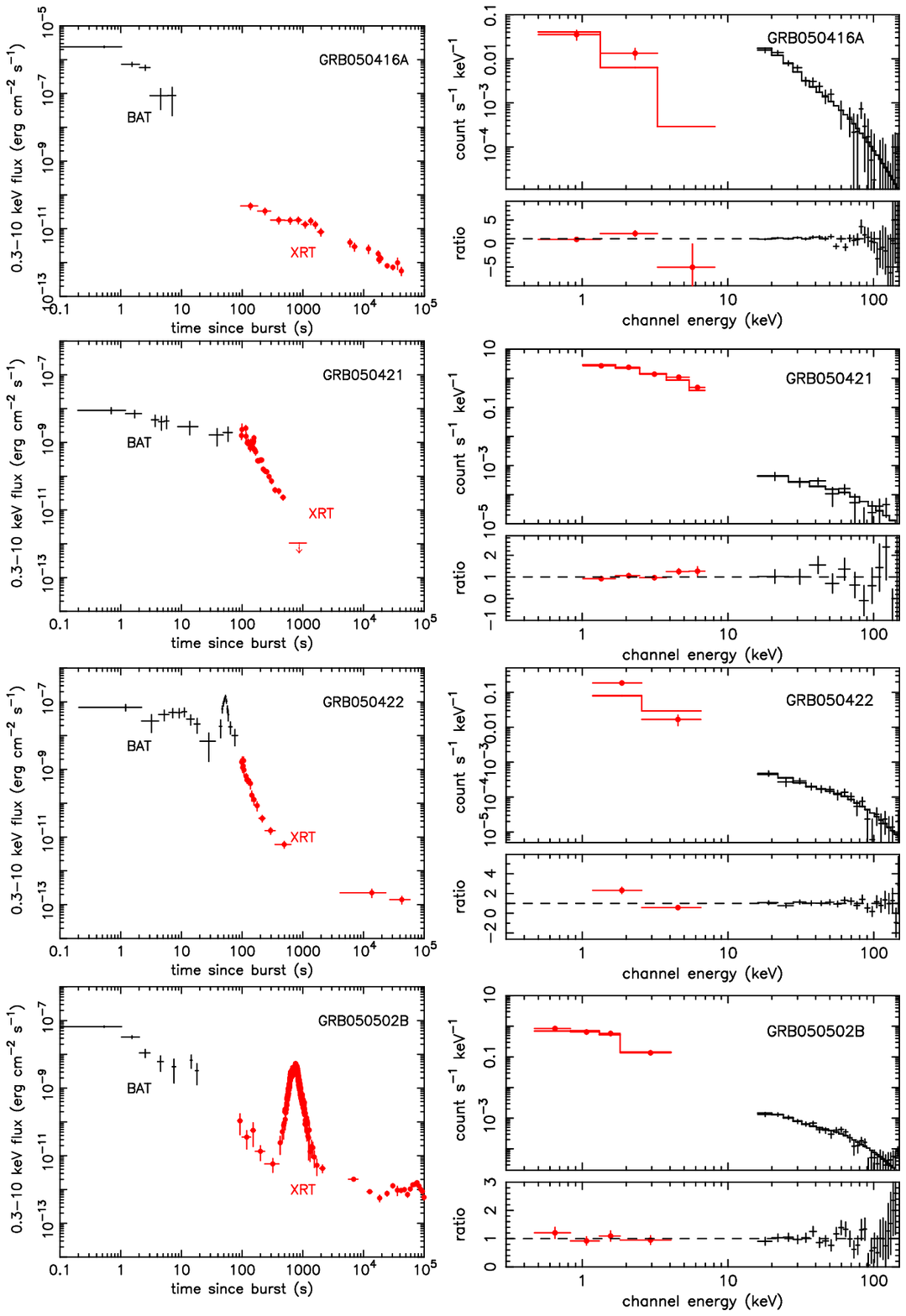}
}
\centerline{Fig. 2. --- continued.}

\clearpage

\centerline{
\includegraphics[]{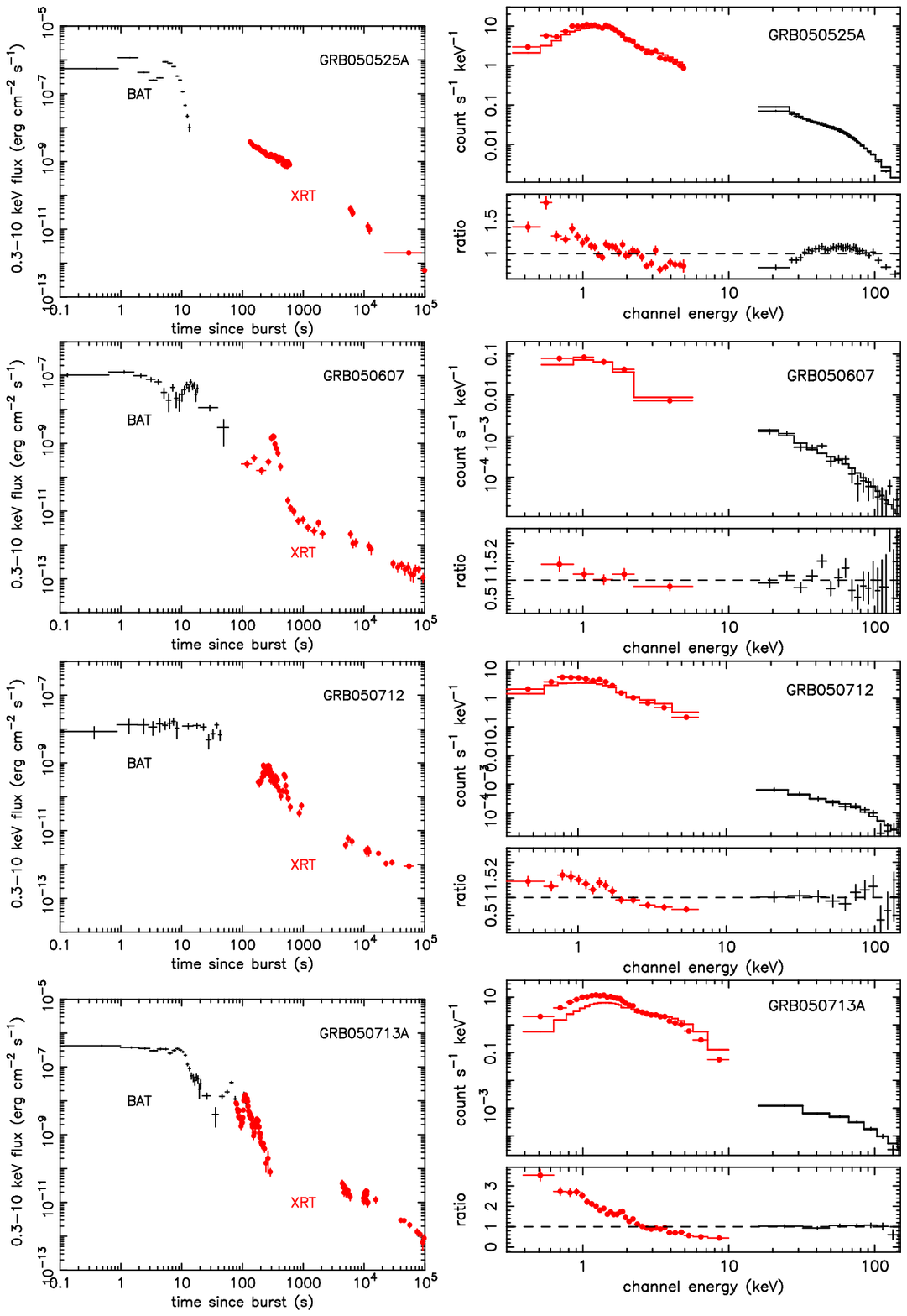}
}
\centerline{Fig. 2. --- continued.}

\clearpage

\centerline{
\includegraphics[]{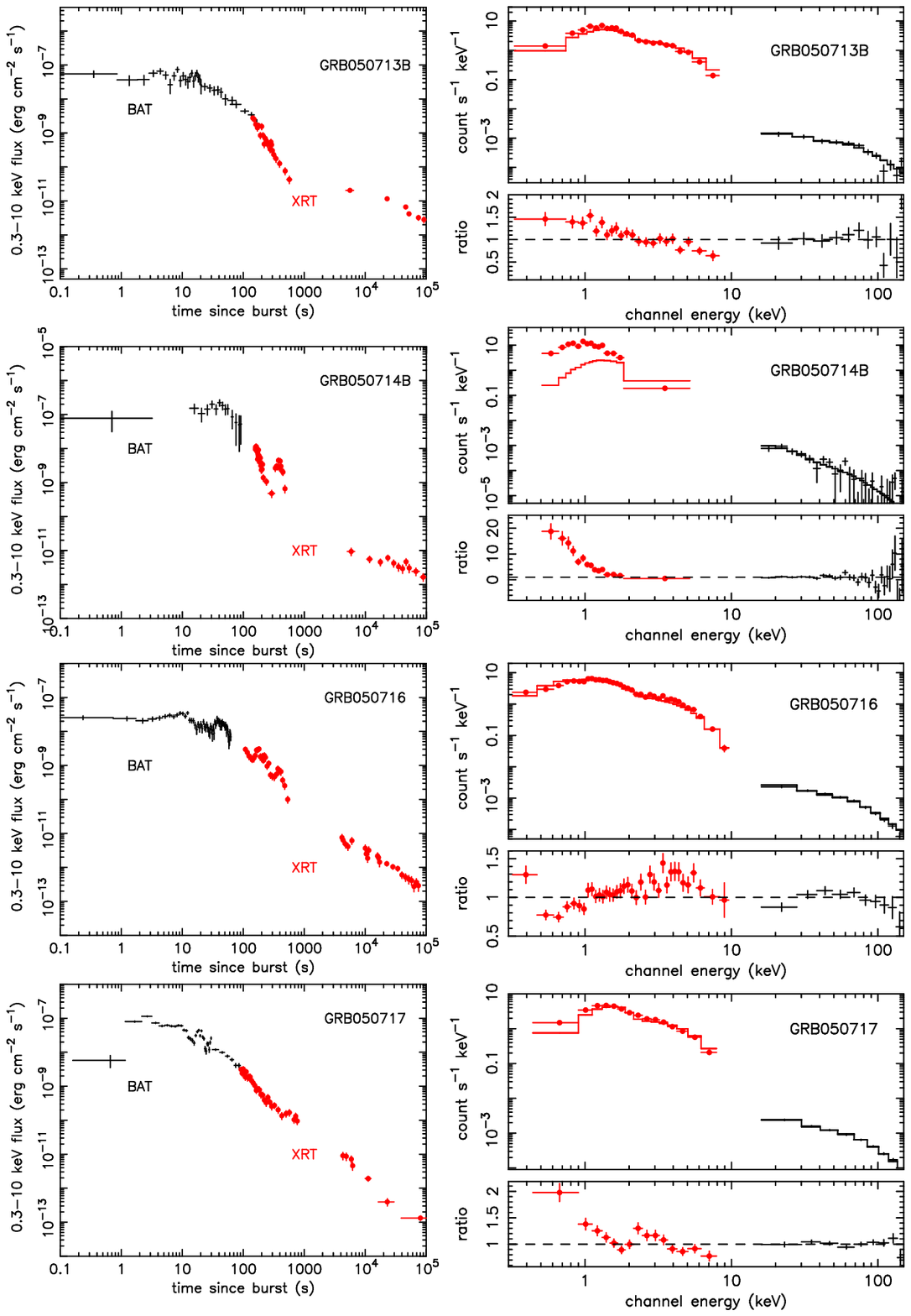}
}
\centerline{Fig. 2. --- continued.}

\clearpage

\centerline{
\includegraphics[]{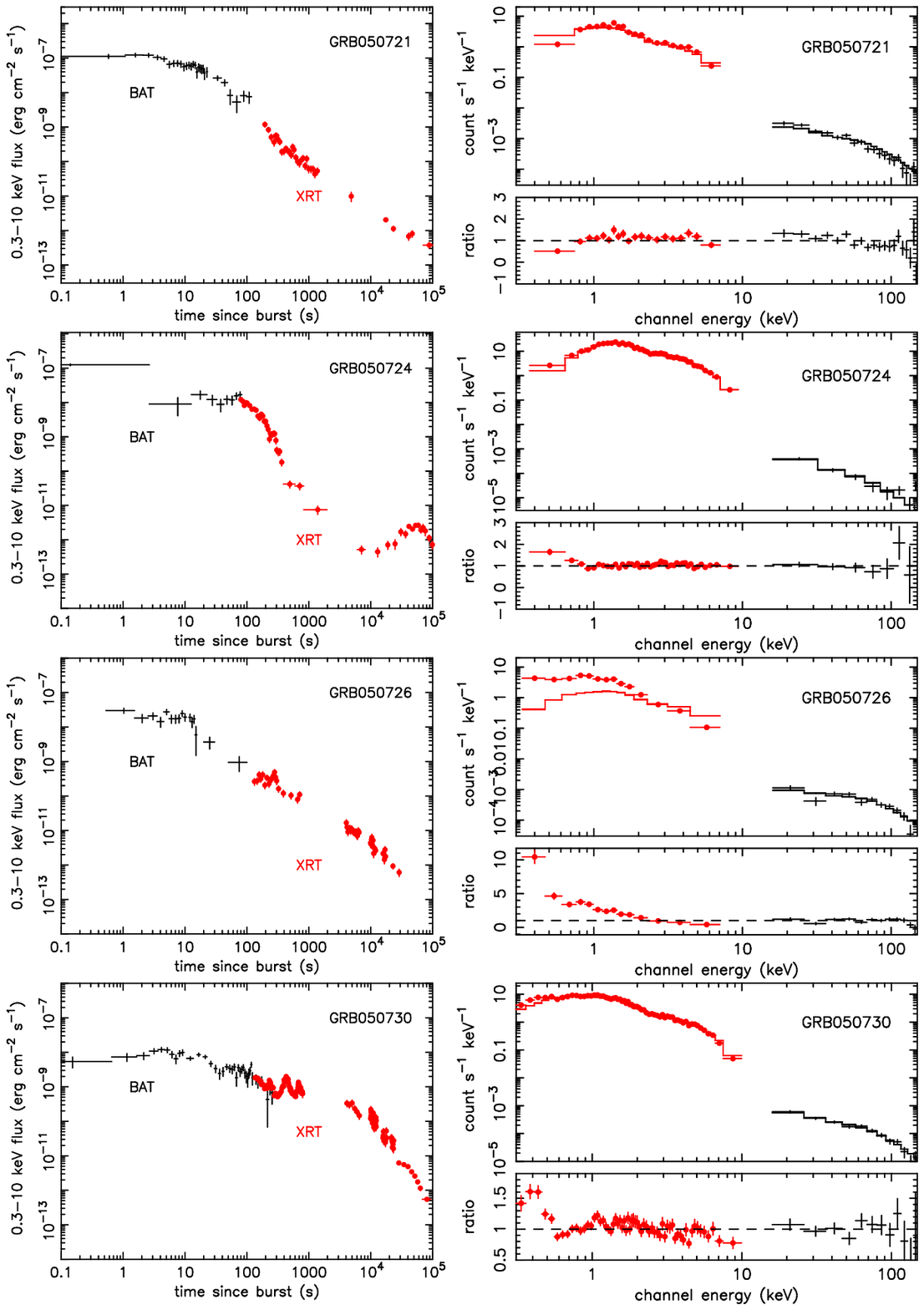}
}
\centerline{Fig. 2. --- continued.}

\clearpage

\centerline{
\includegraphics[]{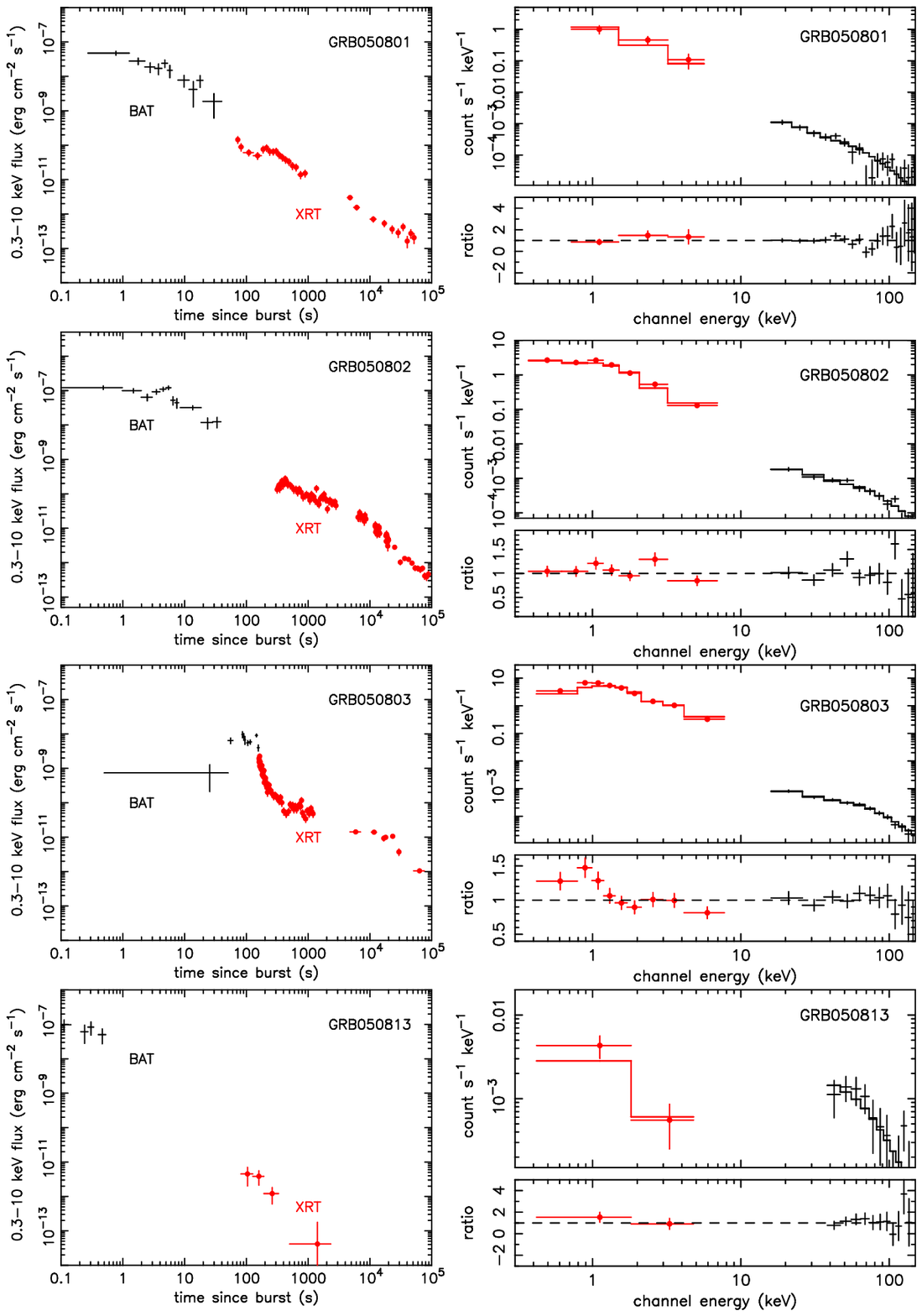}
}
\centerline{Fig. 2. --- continued.}

\clearpage

\centerline{
\includegraphics[]{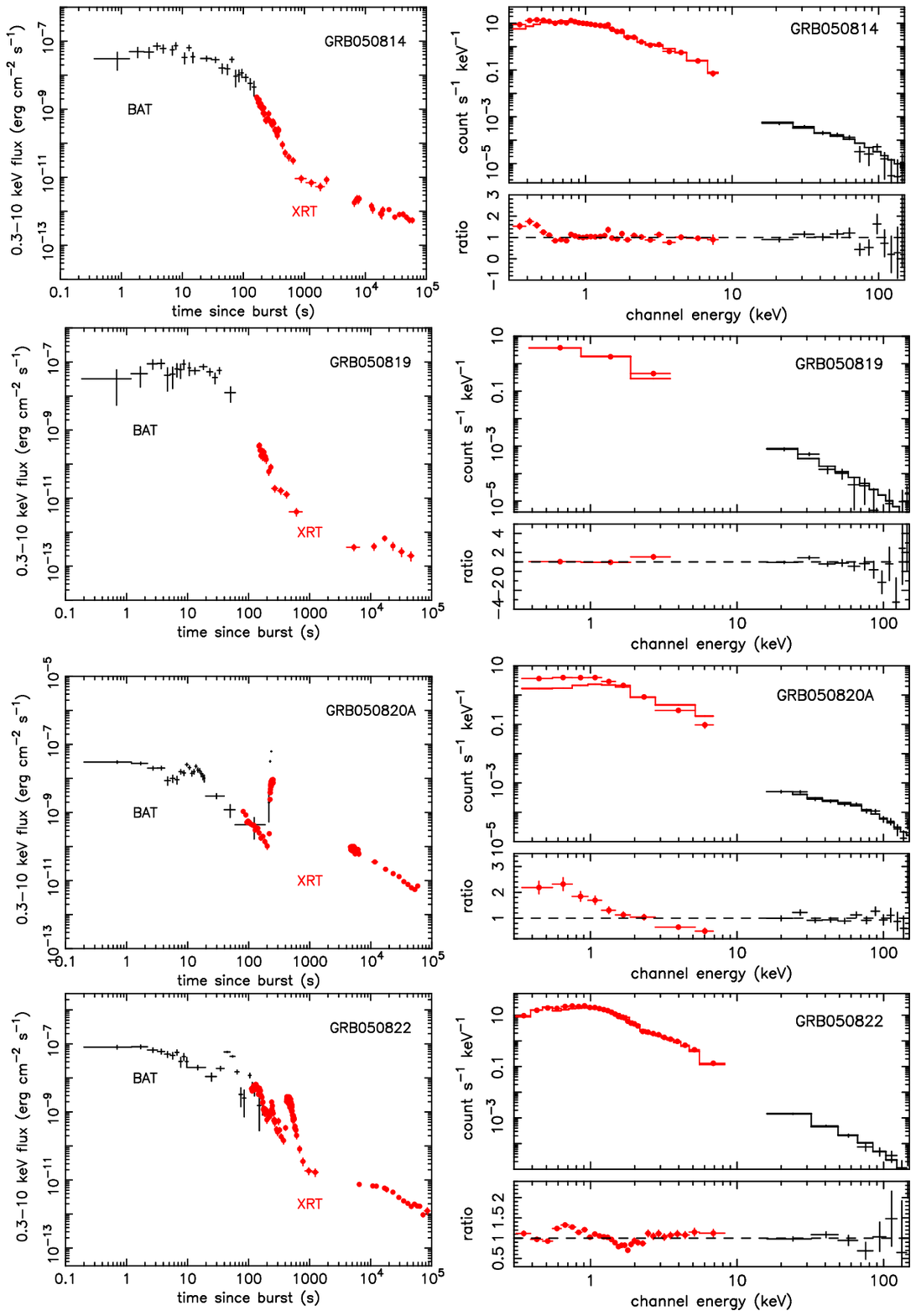}
}
\centerline{Fig. 2. --- continued.}

\clearpage

\centerline{
\includegraphics[]{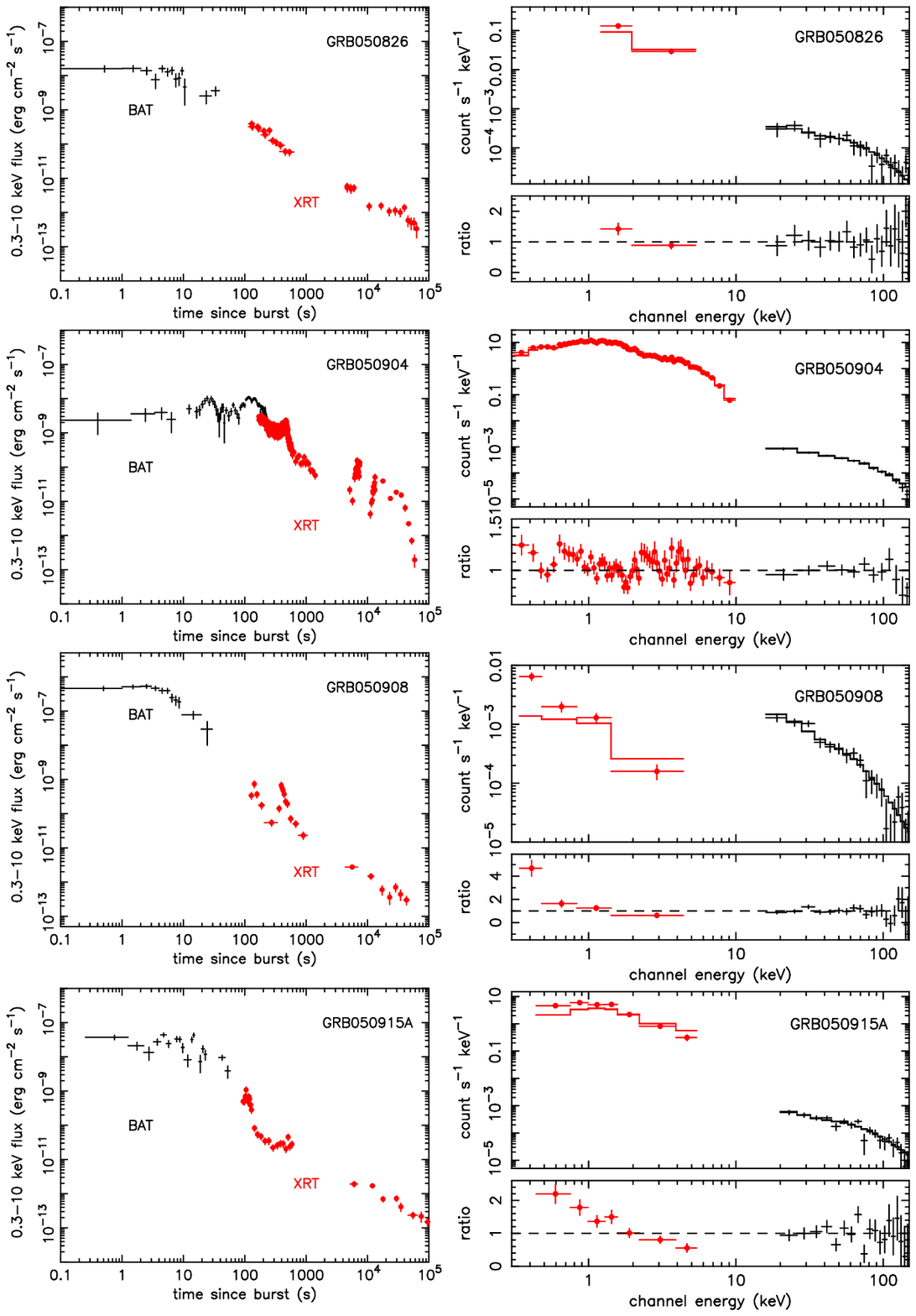}
}
\centerline{Fig. 2. --- continued.}

\clearpage

\centerline{
\includegraphics[]{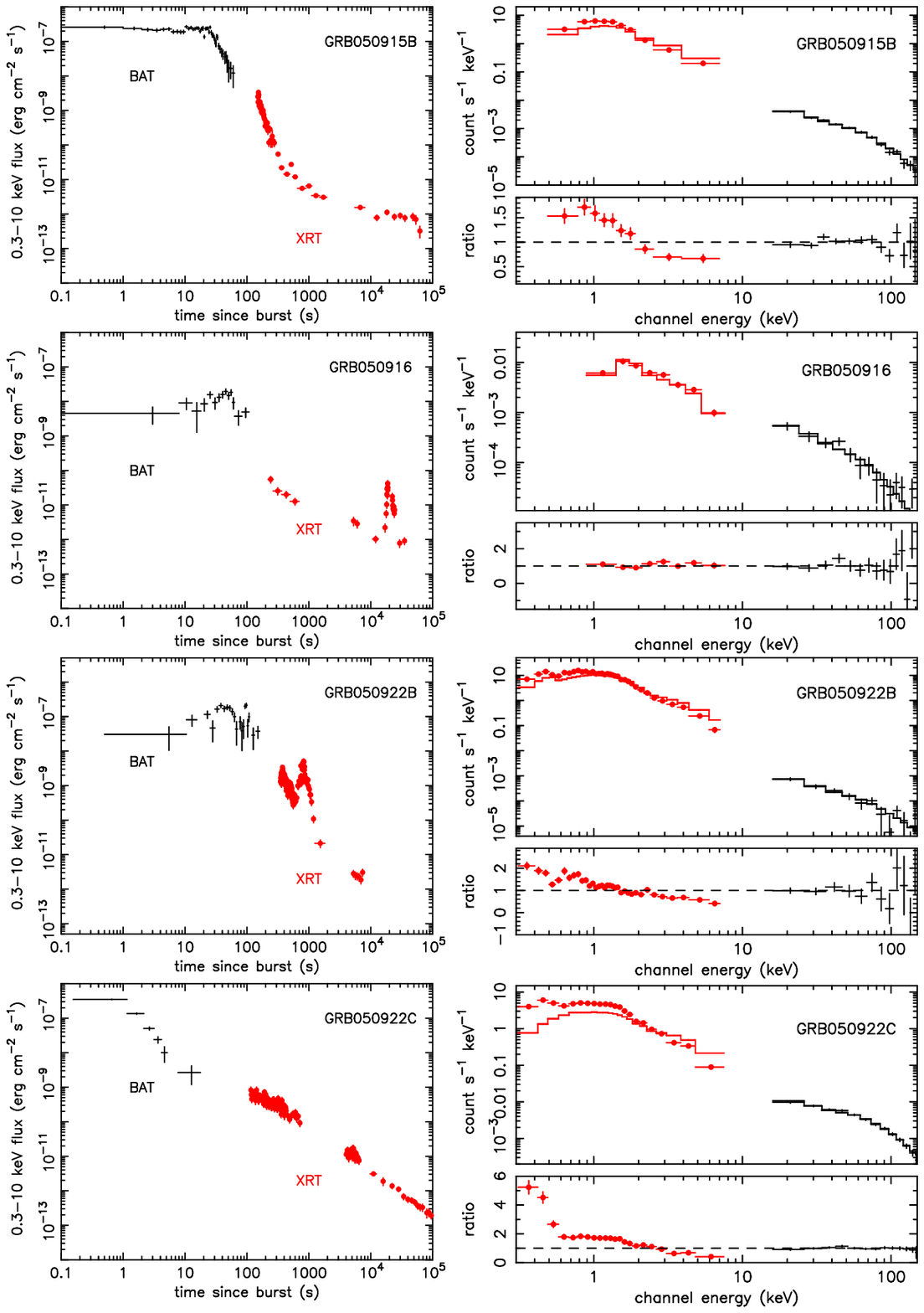}
}
\centerline{Fig. 2. --- continued.}

\clearpage

\clearpage

\clearpage

\clearpage

\clearpage

\clearpage

\clearpage

\clearpage

\clearpage

\clearpage

\clearpage

\clearpage

\end{document}